\documentclass[iop,numberedappendix]{emulateapj}
\usepackage{graphicx}
\usepackage{rotating}
\usepackage{verbatim}
\usepackage{amsmath}
\usepackage{amssymb}
\usepackage{natbib}
\usepackage[colorlinks=true,
            linkcolor=blue,
            urlcolor=blue,
            citecolor=blue]{hyperref}

\slugcomment{2015 ApJ in press}

\shorttitle{}
\shortauthors{CISTERNAS ET AL.}

\begin{document}

\title{The Role of Bars in AGN Fueling in Disk Galaxies Over the Last Seven Billion Years}


\author{Mauricio Cisternas$^{1,2}$,
Kartik Sheth$^{3}$,
Mara Salvato$^{4}$,
Johan H. Knapen$^{1,2}$,
Francesca Civano$^{5,6}$,\\
and
Paola Santini$^{7}$
}
\email{mauricio@iac.es}


\affil{$^{1}$ Instituto de Astrof\'{\i}sica de Canarias, E-38205 La Laguna, Tenerife, Spain}
\affil{$^{2}$ Departamento de Astrof\'{\i}sica, Universidad de La Laguna, E-38205 La Laguna, Tenerife, Spain}
\affil{$^{3}$ National Radio Astronomy Observatory, 520 Edgemont Road, Charlottesville, VA 22903, USA}
\affil{$^{4}$ Max-Planck-Institut f\"ur extraterrestrische Physik, Giessenbachstrasse, D-85748 Garching, Germany}
\affil{$^{5}$ Yale Center for Astronomy and Astrophysics, 260 Whitney Avenue, New Haven, CT 06520, USA}
\affil{$^{6}$ Harvard-Smithsonian Center for Astrophysics, 60 Garden Street, Cambridge, MA 02138, USA}
\affil{$^{7}$ INAF - Osservatorio Astronomico di Roma, via di Frascati 33, I-00040, Monte Porzio Catone, Italy}

\begin{abstract}
We present empirical constraints on the influence of stellar bars on the fueling of active galactic nuclei (AGNs) out to $z=0.84$ using a sample of X-ray-selected AGNs hosted in luminous non-interacting face-on and moderately inclined disk galaxies from the {\em Chandra} COSMOS survey.
Using high-resolution {\em Hubble Space Telescope} imaging to identify bars, we find that the fraction of barred active galaxies displays a similar behavior as that of inactive spirals, declining with redshift from 71\% at $z\sim0.3$, to 35\% at $z\sim0.8$.
With active galaxies being typically massive, we compare them against a mass-matched sample of inactive spirals
and show that, while at face value the AGN bar fraction is slightly higher at all redshifts, we cannot rule out that the bar fractions of active and inactive galaxies are the same.
The presence of a bar has no influence on the AGN strength, with barred and unbarred active galaxies showing equivalent X-ray luminosity distributions.
From our results, we conclude that
the occurrence and the efficiency of the fueling process is independent of the large scale structure of a galaxy.
The role of bars, if any, may be restricted to providing the suitable conditions for black hole fueling to occur, i.e., bring a fresh supply of gas to the central 100 pc.
At the high-redshift end, we find that roughly 60\% of active disk galaxies are unbarred.
We speculate this to be related with the known dynamical state of disks at higher redshifts---more gas-rich and prone to instabilities than local spirals---which could also lead to gas inflows without the need of bars.
\end{abstract}



\defcitealias{sheth08}{S08}

\section{Introduction}

The debate on the nature and the relevance of the different mechanisms able to trigger active galactic nucleus (AGN) activity has been reignited in the last few years. 
Violent events such as major galaxy mergers are an ideal way of bringing a large supply of gas to the innermost region of a galaxy, and support a nuclear starburst and promote supermassive black hole (BH) accretion \citep{sanders88a,barnes&hernquist91}.
Nevertheless, recent studies of AGN host galaxies have established that, at least since $z\sim2$, a significant fraction of BH growth is occurring in seemingly undisturbed disk galaxies  \citep{gabor09,georgakakis09,cisternas11a,schawinski11,kocevski12},
which appear to share the same evolutionary path as normal ``main sequence" star forming galaxies \citep{mullaney12a,rosario13,goulding14}.

These findings suggest that the role of major mergers may be confined to very specific regions of the AGN parameter space, (e.g., the highest luminosities, \citealt{canalizo&stockton01,ramosalmeida12,treister12}), and alternative mechanisms may be more important than previously thought.
Recent models have started to disentangle the relevance of different fueling mechanisms, acknowledging that major mergers alone cannot account for the observed AGN luminosity function, and an additional significant contribution from a secular, i.e., less violent, AGN triggering model is required \citep{draper12,hopkins&kocevski14,menci14}.

Among the possible secular mechanisms that could remove angular momentum from the gaseous interstellar medium, and help to bring it down to the central regions of the galaxy, stellar bars have received a great deal of attention.
They are highly common in the nearby universe, with nearly two out of three galaxies having a bar \citep[e.g.,][]{devaucouleurs63,eskridge00,knapen00,menendez07,marinova&jogee07}.
The non-axisymmetric potential of a bar is able to exert torques which result in radial gaseous inflows toward the galactic center, leading to high central concentrations of molecular gas and nuclear star formation \citep{athanassoula92b, martin95, ho97b,sakamoto99,regan99b,sheth05,ellison11,coelho11}.
Once the gas is accumulated within the central hundred pc, further dynamical instabilities could potentially transport the gas inward and feed a BH \citep[e.g., the ``bars within bars" picture,][]{shlosman89}.

A number of studies have investigated a tentative ``bar-AGN connection" in the local universe with conflicting results.
While Seyfert galaxies tend to have a higher fraction of large-scale bars with respect to ``normal" spirals \citep{knapen00,laine02}, neither the presence nor the strength of a bar have an effect on the level of ongoing AGN activity \citep{ho97b, cisternas13}.
Moving to smaller scales, observational programs targeting the central regions of nearby galaxies with the {\em Hubble Space Telescope (HST),} have found nuclear bars on a minority of active galaxies \citep{regan99a,martini01}.
These studies, however, revealed dust spiral structure connecting to the large-scale bar and extending all the way down to the unresolved nucleus.
While these nuclear structures are a promising mechanism to transport material down to sub-pc scales, dust spirals in various configurations are found in both, active and inactive galaxies with comparable frequency \citep{martini03b}.
Still, there appear to be systematic morphological differences in the central structure of starburst, Seyfert, and LINER galaxies, hinting at a possible evolutionary sequence \citep{hunt&malkan04}.
From the point of view of the molecular gas at $\sim$100 pc scales, a variety of morphologies and kinematic modes are found in nearby active galaxies, hinting at a hierarchy of angular momentum transport mechanisms over different spatial scales \citep{garcia-burillo05,haan09}.

Stellar bars are indeed efficient at bringing large supplies of gas to the central regions of a galaxy, but catching BH fueling in the act is challenging given the timescales involved, with an episode of AGN activity  typically lasting just a few million years \citep[e.g.,][]{martini04_lifetime}.
Some barred galaxies show a gas deficit within the bar region, and thus it is likely their central gas supply was already converted into stars \citep{sheth05}, and any episode of significant BH accretion would have already occurred.
Additionally, the fact that AGN activity is also observed in unbarred galaxies means that bars are not a necessary condition for AGN fueling to occur, and further complicate the detection of direct connections.
Therefore, the limited success of past studies does not necessarily mean that bars do not play a role in promoting BH growth in disk galaxies.
Regardless of the results to date, all previous studies have concentrated on nearby galaxies and no empirical constraints on the role of bars on AGN fueling exist beyond $z\sim0$, where higher luminosity AGNs are more common, and the fueling requirements could be different than those from local Seyferts.

Motivated by
(1) the lack of observational results at intermediate-high redshifts,
(2) the widespread ongoing AGN activity in disk galaxies, 
(3) the decrease of the fraction of barred galaxies with cosmic time (\citealt{sheth08}, hereafter \citetalias{sheth08}),
and (4) the increase in the number density of luminous AGNs with redshift \citep{ueda14},
in this Paper
we establish the actual relevance of bars in the fueling of nuclear activity as a function of redshift.
We assemble a sample of 95 X-ray selected AGNs  from the {\em Chandra} COSMOS survey \citep[C-COSMOS;][]{cosmos_chandra1} hosted in luminous face-on and moderately inclined disk galaxies over the redshift range $0.15<z<0.84$.
With such a sample, we are able to directly compare the levels of nuclear activity in barred and unbarred galaxies and, for the first time, study the evolution of the fraction of bars in active galaxies over the last seven billion years.


\section{Dataset and Sample Selection}

We build our sample of active disk galaxies by selecting AGNs based on their X-ray emission, and subsequently using their multiwavelength photometry and optical morphologies to establish the nature of their host galaxies.

One of the primary goals of this study is to perform a direct comparison of the bar fraction of active disk galaxies against the well-established evolving bar fraction of inactive galaxies presented by \citetalias{sheth08}.
In brief, \citetalias{sheth08} measured the evolution of the bar fraction over $0.1<z<0.8$ using over two thousand luminous spiral galaxies from the COSMOS field \citep{cosmos}.
In order to build their sample of disk galaxies, they required
(1) an upper redshift cut to stay in the optical regime,
(2) a redshift-dependent galaxy luminosity cut,
(3) removal of elliptical and lenticular galaxies, as well as clearly interacting and distorted galaxies,
and (4) galaxies with moderate disk inclinations, to allow an unambiguous identification of bars.
Therefore, to recover a comparable sample of disk galaxies, below we describe the application of each of their selection criteria to our initial sample of AGNs.

The COSMOS field has been observed with both {\em XMM-Newton} and {\em Chandra}.
While the {\em XMM-Newton} observations comprise the whole 2 deg$^2$ field, the C-COSMOS field consists of only the central 0.9 deg$^2$.
Because of its lower flux limits (3 times below that of {\em XMM-Newton}), however, {\em Chandra} allows us to recover additional lower-luminosity AGNs otherwise missed by {\em XMM-Newton}.
We therefore opt to use the {\em Chandra} observations: the tradeoff in smaller area for better sensitivity is the appropriate choice considering these additional lower-luminosity AGNs are likely to be found in spiral galaxies.

\begin{figure*}[t]
\begin{center}
\includegraphics[scale=0.51]{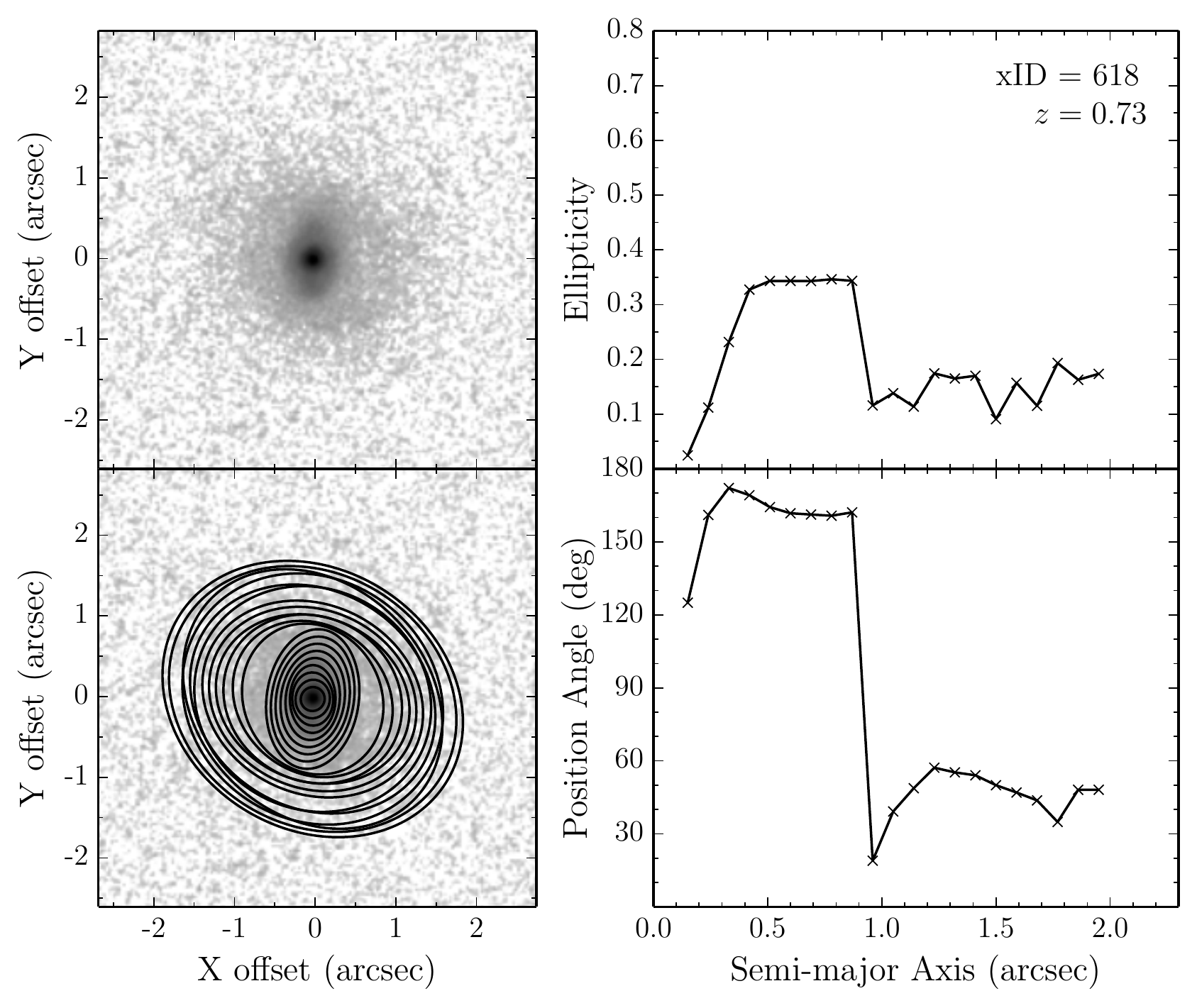}
\includegraphics[scale=0.51]{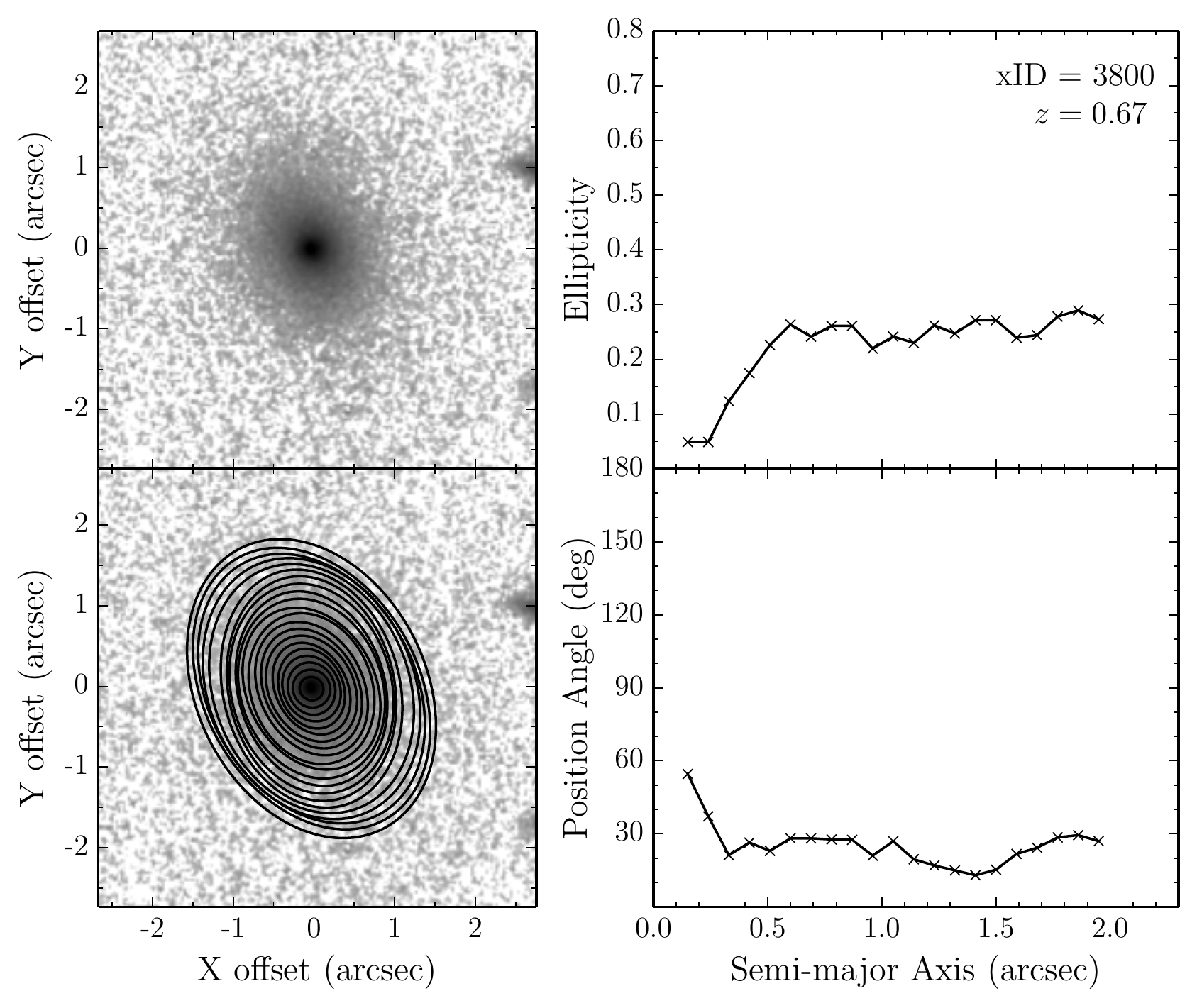}
\caption{Example results from the ellipse fitting task for a barred active galaxy at $z=0.73$ (left panels), and for an unbarred active galaxy at $z=0.67$ corresponding to the X-ray sources xID=618 and xID=3800 respectively \citep{cosmos_civano12}.
For each galaxy, we show the {\em HST}/ACS F814W cutout (top-left), as well as the galaxy isophotes overlaid (bottom-left).
For the barred galaxy, its ellipticity profile (top-right) shows a steep drop at 0\farcs9, which is matched by a sharp change in its position angle profile (bottom-right) at the same semi-major axis distance, in contrast with the rather smooth profiles of the unbarred galaxy.\label{fig_ellipse}}
\end{center}
\end{figure*}

From the C-COSMOS catalog of 1761 X-ray point sources with optical counterparts \citep{cosmos_civano12}, we perform an initial selection of 357 sources with $z<0.835$.
Spectroscopic redshifts are available for 80\% of the sample, while for the rest, we use photometric redshifts from \citet{cosmos_salvato11}, carefully determined to account for the AGN contribution and optical variability, and with an accuracy of $\sigma\sim0.015$.
Given that we will identify bars using the {\em HST}/ACS images in the F814W (broad $I$) filter, the upper redshift boundary is determined by the bandpass shifting in this band.
As thoroughly discussed in \citetalias{sheth08}, bars tend to disappear as one goes blueward of the rest frame $u$-band, and in order to remain consistent with the \citetalias{sheth08} sample selection we impose an upper redshift limit of $z=0.835$ in order to stay in the optical.
We also impose a lower redshift cut at $z=0.15$, because of (1) the low number of X-ray sources below these redshifts, and (2) their typically modest X-ray luminosities (see below), which casts doubt on their AGN nature.
We retrieve the latest (v2.0) processed {\em HST}/ACS images \citep{cosmos_acs}, with a pixel scale of 0\farcs03/pixel and a resolution of 0\farcs1, corresponding to $\sim$0.8 kpc at $z=0.84$.

Following \citetalias{sheth08}, we select galaxies brighter than $L^{\ast}_{\rm V}$, the ``knee'' of the luminosity function.
Because of the significant redshift interval spanned by our sample, we account for luminosity evolution in the $V$-band luminosity function \citep{ilbert05} in order to remain consistent at all redshifts.
Physical properties of the galaxies and X-ray-detected AGNs
are available from \citet{cosmos_ilbert09} and \citet{cosmos_salvato11} respectively, through
spectral energy distribution (SED) fitting to the vast multiwavelength data available in COSMOS.
Using the shape of the SEDs, we remove elliptical and S0 galaxies from the sample \citep[corresponding to templates 1 to 6, see][]{cosmos_ilbert09}.

The next step in our sample culling consists in visually identifying and removing irregular and interacting galaxies, as well as point-like objects, as confirmed by the original {\em HST} source catalog \citep{cosmos_leauthaud07}.
In order to identify bars accurately, we impose an inclination cut ($i<65^{\circ}$) measured from the axial ratio of the galactic disk, which in turn is determined
using the latest version of GALFIT \citep{galfit10} to perform a two-dimensional image decomposition.
To each galaxy, we fit a model made up of two components: an exponential profile to account for the disk, and a S\'ersic profile to represent the bulge.

Finally, to ensure that our sample of X-ray-selected AGNs is not contaminated by purely star-forming galaxies without active nuclei, we impose a lower cut at $L_{\rm X}=10^{42}$ erg s$^{-1}$, where $L_{\rm X}$ corresponds to the intrinsic X-ray luminosity in the hard 2--10 keV band.
This boundary in $L_{\rm X}$ is a rather conservative upper limit to the maximum X-ray emission we would expect from pure star formation, given that typical X-ray luminosities from local luminous infrared galaxies, i.e., the most strongly star-forming systems, are always below this value \citep[e.g.,][]{lira02,iwasawa09, lehmer10}.
Intrinsic X-ray luminosities are computed from the observed fluxes \citep{cosmos_civano12}, correcting for absorption using the band ratio, and applying a $k$-correction assuming a power-law photon index $\Gamma=1.4$.
This removes an additional 10 galaxies, particularly at the low-redshift end, and leaves our final sample size at 95 AGNs hosted in face-on and moderately inclined, $L^{\ast}$ disk galaxies.


\section{Bar Identification}

Bars are identified by visually examining the {\em HST}/ACS images as well as the ellipticity and position angle (P.A.) profiles.
As is customary in studies of barred galaxies (e.g., \citealt{knapen00,sheth00,laine02,menendez07}, \citetalias{sheth08}), these measurements were obtained by fitting ellipses to the galaxy isophotes, using the IRAF task \verb"ellipse".
Unbarred disk galaxy isophotes tend to show a smooth rise in ellipticity with radius as they go from the round bulge-region toward the disk.
On the other hand, the ellipticity profile in barred galaxies is characterized by a monotonic increment along the bar region followed by a rather sharp drop once the bar ends and the isophotes start tracing the galactic disk.
Ideally, barred galaxies will also display a constant P.A. along the bar region, at the end of which a sudden change indicates the transition to disk isophotes, due to different P.A.s between bar and disk.
An example of these features is presented in Figure \ref{fig_ellipse} for a barred active galaxy at the high redshift end of our sample, shown together with an unbarred galaxy for comparison.

With the aim of removing subjectivity in the process of determining whether a galaxy is barred or not, some studies opt to use some hard cuts on the variations of ellipticity $\Delta\epsilon$ and $\Delta$P.A.
There can be cases, however, in which this signature is not entirely clear due to e.g., the alignment of the P.A.s of bar and disk, or the ellipticity drop signature being produced by spiral arms rather than an actual stellar bar\footnote{In addition, a warped stellar disk may also introduce changes in the ellipticity and P.A. profiles, yet its effect tends to be more relevant toward the outskirts of the disk, way beyond the end of the bar.} \citep[for a detailed discussion, see][]{menendez07}.
The above shows the importance of complementing the ellipse-based bar identification with an additional independent assessment of the bar presence, such as visual inspection of the galaxy images, to ensure that the bar signature is not being masked by the aforementioned sources of confusion.
Considering the relatively manageable sample size of the present study, we also opt to visually analyze each individual set of galaxy images, isophotes, and ellipticity and P.A. profiles to identify barred galaxies.

From our analysis, we identify 47 out of 95 active galaxies as barred.
In the following section, we dissect in detail the bar fraction as a function of redshift, and study how it compares to that of inactive spiral galaxies.


\section{Results \& Discussion}

\subsection{Evolution of the AGN Bar Fraction}

\begin{figure}[t]
\begin{center}
\includegraphics[scale=0.65]{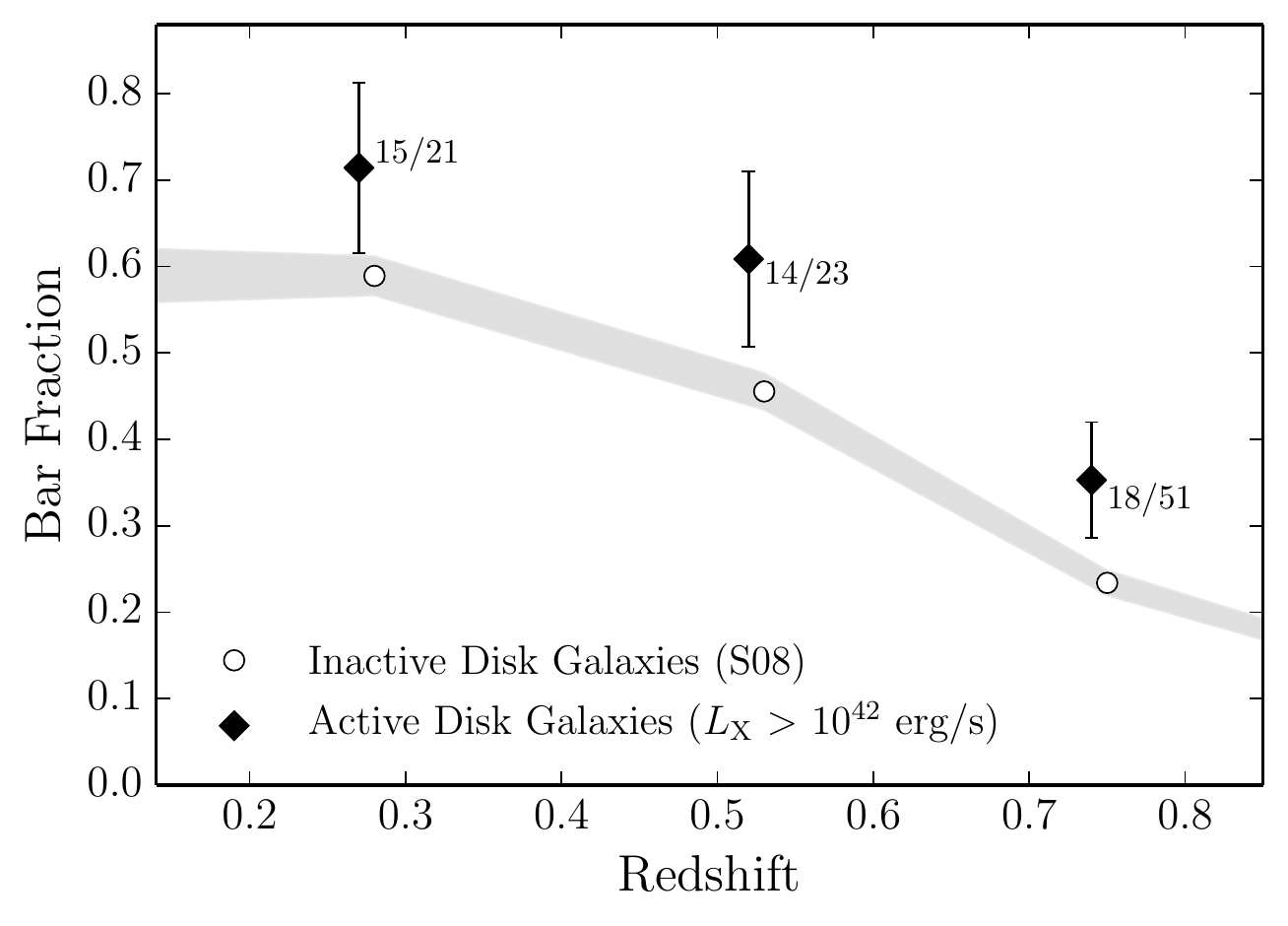}
\caption{Bar fraction of active galaxies as a function of redshift (filled diamonds) evaluated over three redshift bins: 0.15--0.40, 0.40--0.65, and 0.65--0.84, with the fraction itself indicated.
We also show the bar fraction of inactive disk galaxies (empty circles) in the COSMOS field computed for the same redshift bins, and extrapolated to the expected values at $z=0$ and $z=1$, based on a local disk sample \citepalias{sheth08} and simulations \citep{kraljic12} respectively.
Vertical error bars, and shaded area represent the standard errors ($1\sigma$) assuming binomial statistics.\label{fig_bf}}
\end{center}
\end{figure}

The bar fraction, $f_{\rm bar}$, is defined as the number of barred disk galaxies over the total number of disk galaxies in a given sample.
For inactive galaxies, $f_{\rm bar}$ is not constant but evolves with redshift, going from $\sim$0.6 at $z=0$ to $\sim$0.2 at $z=0.8$ (\citetalias{sheth08}, \citealt{cameron10,kraljic12,melvin14}).
Because of this strong evolution over the last 7 Gyr, it is of great interest to quantify not just the global fraction of active disk galaxies, but also its behavior with redshift.
In Figure \ref{fig_bf} we present the first determination of $f_{\rm bar}$ for active $L^{\ast}$ disk galaxies as a function of redshift.
Uncertainties ($1\sigma$) in the bar fraction, shown as vertical error bars, are calculated from binomial statistics, i.e., for a given bar fraction $f$, and subsample size $N$, the standard error is determined by $\sqrt{f(1-f)/N}$.

To put these new findings in the broader context of galaxy evolution, we compare it directly with the bar fraction of inactive galaxies built from the S08 sample.
It is important to mention that comparisons of results from different studies can be tricky because of several subtle differences in the sample selection process.
In our particular case, however, this comparison is valid since as described in Sections 2 and 3, the selection of the sample of active disks as well as the bar identification was carefully designed to reproduce that of \citetalias{sheth08}.

Since the publication of \citetalias{sheth08}, the addition of photometric bands to the COSMOS catalog over time has permitted a continuous refinement of the photometric redshifts and physical properties derived for the galaxy sample.
Therefore, we updated the \citetalias{sheth08} sample with the latest photometric redshifts and absolute $V$-band magnitudes available from \citet{cosmos_ilbert09}.
Additionally, we removed X-ray-detected sources, which yields a final sample of 1651 $L^{\ast}$ disk galaxies in the redshift range $0.15<z<0.84$.
With the updated sample, in Figure \ref{fig_bf} we present the bar fraction of inactive galaxies calculated over the same redshift bins, with uncertainties represented by the shaded area.
For illustrative purposes, the inactive bar fraction is shown extrapolated to the expected values at $z=0$ of $f_{bar}=0.59$ and $z=1$ of $f_{bar}=0.10$, based on a local disk sample \citepalias{sheth08} and simulations \citep{kraljic12} respectively.
Details on the bar fractions of active and inactive galaxies are given in Table \ref{tbl_1}.

We find that the bar fraction of active $L^{\ast}$ disk galaxies declines with redshift, dropping over a factor of 2 from $z\sim0.25$ to $z\sim0.75$, and showing a similar trend as the bar fraction of ``normal" spirals known from the literature. 
When directly compared to the bar fraction of inactive galaxies, our results show that, at all redshifts probed, active disk galaxies are barred more frequently than inactive galaxies beyond uncertainties.
At face value, this result could suggest that active and inactive disk galaxies at $0.15<z<0.84$ are structurally different, and the presence of a stellar bar could be strongly related to the occurrence of AGN activity.
However, as we argue below, much of the difference in the higher redshift bins is due to selection effects.

\begin{figure*}[ht]
\begin{center}
\includegraphics[scale=0.67]{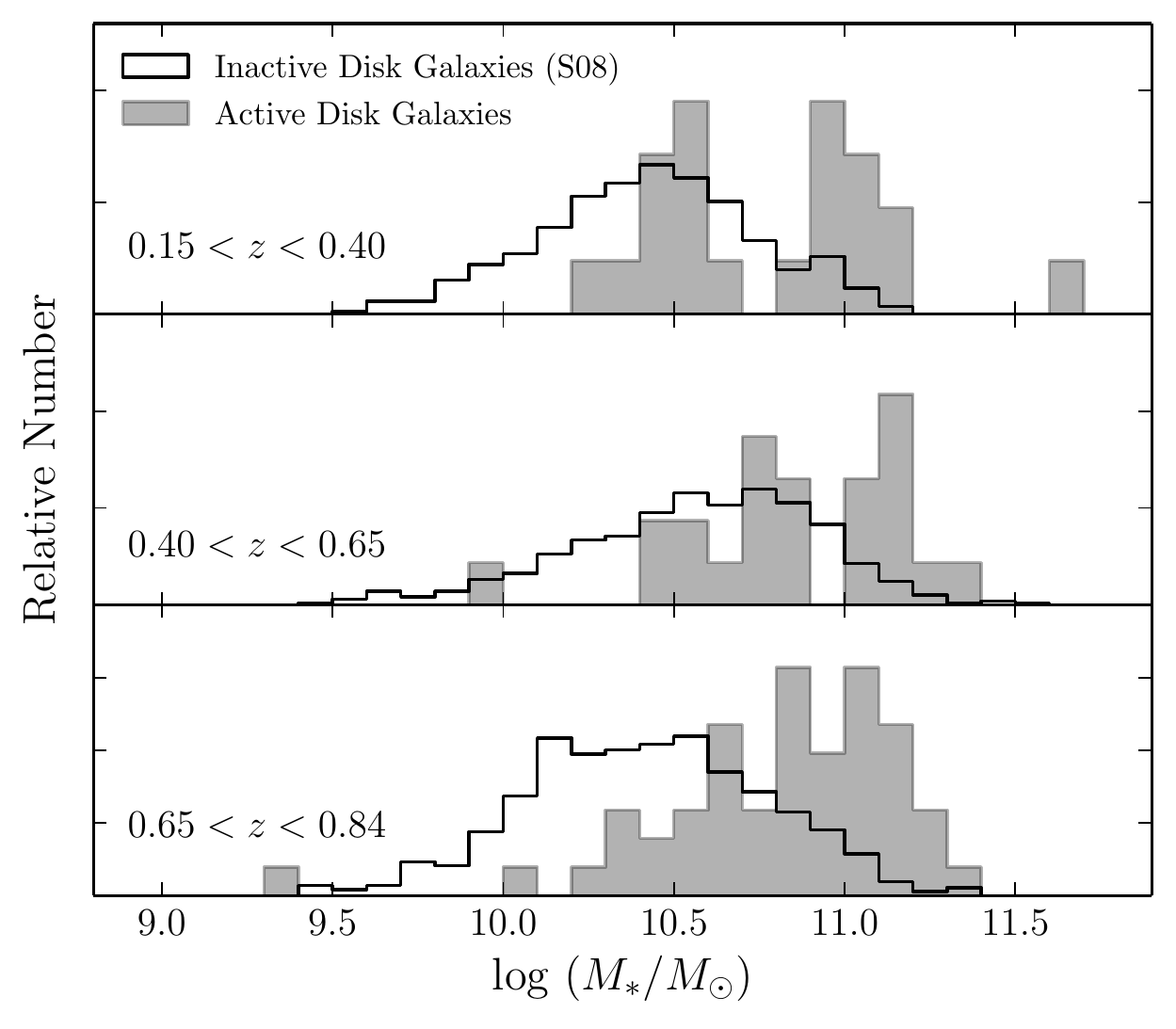}
\includegraphics[scale=0.67]{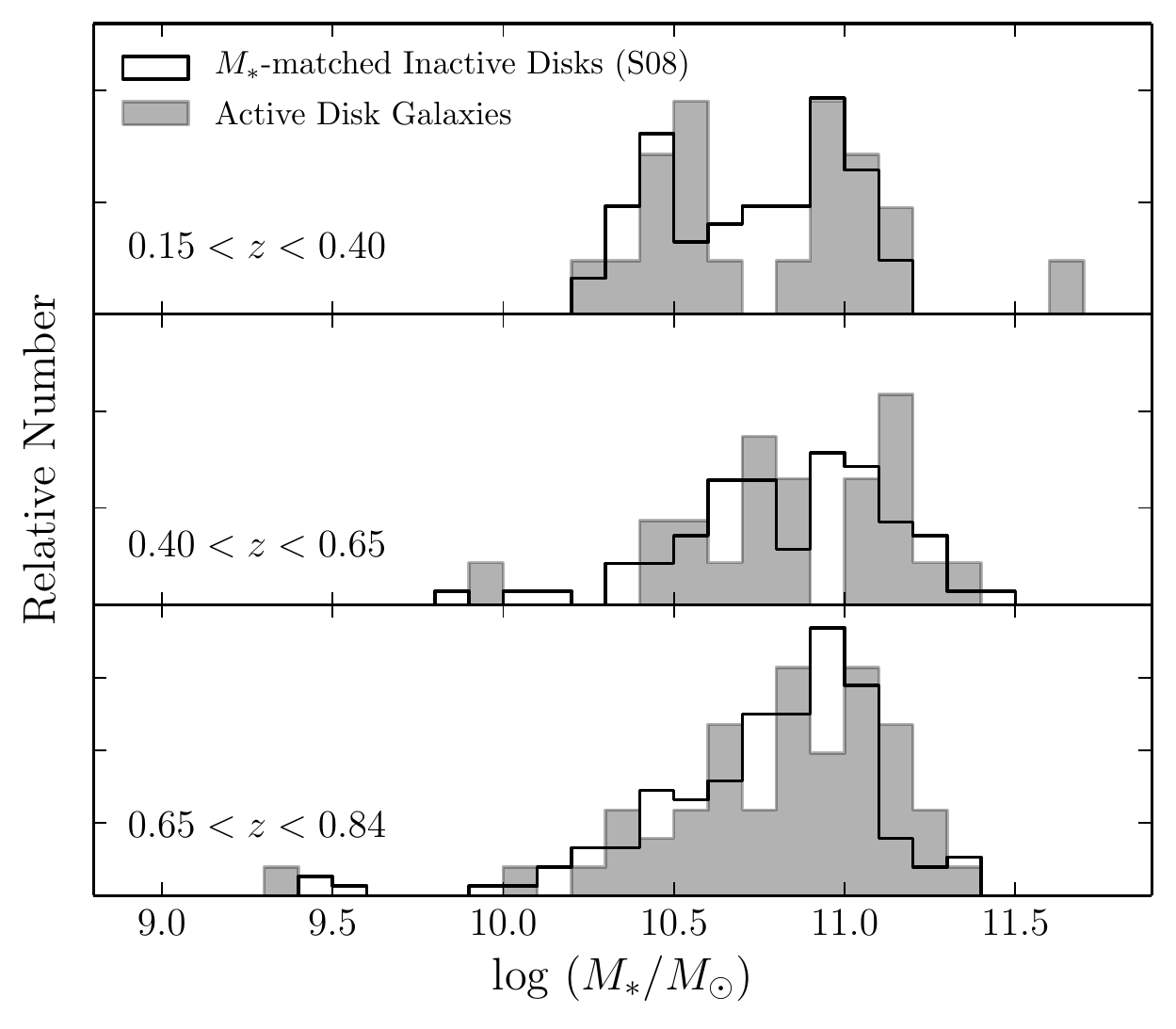}
\caption{Normalized stellar mass distributions of the AGN disk sample (gray histograms) compared against the whole sample of inactive spirals from the \citetalias{sheth08} sample (left), and against a stellar-mass-matched sample of inactive galaxies built from the \citetalias{sheth08} sample (right) for three redshift bins.
\label{fig_mshist}}
\end{center}
\end{figure*}

\subsubsection{The Stellar Mass Dependence}

Before assessing the significance of the above results, one needs to consider other factors that could influence the bar fraction.
For instance, we know from \citetalias{sheth08} that the bar fraction is stellar-mass-dependent:
the most massive disk galaxies tend to have a higher fraction of bars at all redshifts, and show only a mild evolution with cosmic time.
While the mass-dependence of the bar fraction seems to be stronger at higher redshifts, it  shows a nearly flat behavior at the low-redshift end.

\begin{deluxetable}{lccc}
\tabletypesize{\scriptsize}
\tablecaption{Galaxy Samples and Bar Fractions\label{tbl_1}}
\tablewidth{0.45\textwidth}
\tablehead{
Redshift range & \colhead{$N_{\rm tot}$\tablenotemark{a}} & \colhead{$N_{\rm bar}$\tablenotemark{b}} & \colhead{$f_{\rm bar}$\tablenotemark{c}} }
\startdata
Active\\
0.15--0.40  &  21 & 15 & $0.71\pm0.10$  \\
0.40--0.65  &  23 & 14 & $0.61\pm0.10$  \\
0.65--0.84 &  51 & 18 & $0.35\pm0.07$ \\ \hline
Inactive\\
0.15--0.40  &  426 & 251 & $0.59\pm0.02$  \\
0.40--0.65  &  494 & 225 & $0.46\pm0.02$  \\
0.65--0.84 &  731 & 171 & $0.23\pm0.02$ \\ \hline
Inactive ($M_{\ast}$-matched)\tablenotemark{d}\\
0.15--0.40  &  63 & 36.4 & $0.58\pm0.05$  \\
0.40--0.65  &  69 & 36.5 & $0.53\pm0.05$  \\
0.65--0.84 &  153 & 50.5 & $0.33\pm0.03$
\enddata

\tablenotetext{a}{Total number of galaxies in bin.}
\tablenotetext{b}{Number of barred galaxies.}
\tablenotetext{c}{$f_{\rm bar}=N_{\rm bar}/N_{\rm tot}$, standard error given by $\sqrt{f_{\rm bar}(1-f_{\rm bar})/N_{\rm tot}}$.}
\tablenotetext{d}{Quoted numbers for the control sample correspond to the mean of one thousand realizations.}
\end{deluxetable}

It is therefore relevant to investigate whether the enhancement in the bar fraction of active galaxies observed in Figure \ref{fig_bf} could be due to AGNs typically being hosted by more massive galaxies.
Stellar masses for our samples are available through galaxy template fitting to the observed SEDs by \citet{cosmos_ilbert09}.
Even though the AGN contribution was not taken into account in these estimates, the moderate luminosity of our sample implies a modest contribution to the overall SED, and would suggest that the resulting stellar mass values should not be particularly off.
In order to be more precise, we can account for the AGN contribution by taking the stellar mass estimates from \citet{santini12}, who computed stellar masses to the brighter {\em XMM-Newton} COSMOS sample of active galaxies by fitting combined galaxy and AGN templates.
Fifty-six of our active galaxies are covered by their sample, particularly at the bright end, which should be the most affected by the non-thermal emission.
By comparing the stellar mass estimates of \citet{cosmos_ilbert09} and \citet{santini12} for 47 AGNs in common between both samples, we find a median offset of just 0.1 dex, with the latter estimates being on the massive side.
We therefore opt to use the combined SED (galaxy+AGN) stellar mass estimates when available, and the single galaxy template estimates for the remaining 39 active galaxies.

In the left panels of Figure \ref{fig_mshist}, we show the stellar mass distributions of the AGN host galaxies together with the inactive spirals from \citetalias{sheth08}.
One can immediately observe that indeed, AGN host galaxies are consistently more massive than the inactive field population of spirals, with this effect being more severe in the highest redshift bin.
This clear difference in stellar mass could be the reason for the systematic offset between bar fractions observed in Figure \ref{fig_bf}, as we will indeed show below.

In order to properly compare the bar fractions of active and inactive disks, we have to build a mass-matched control sample of inactive disk galaxies.
To do this, for each AGN we randomly select 3 unique galaxies from the \citetalias{sheth08} sample in the same redshift bin, and within a factor of 1.6 in stellar mass ($\Delta$log$ M_{\ast}/M_{\odot}$=0.2).
For one case, the most massive active galaxy in the lowest redshift bin, not enough control galaxies were found to comply with the criteria, and therefore the search interval in stellar mass was interactively increased by 10\% until $\Delta$log$ M_{\ast}/M_{\odot}$=0.6, when enough matching galaxies were found.
In the right panels of Figure \ref{fig_mshist} we show the resulting stellar mass distributions of the AGN hosts and of the mass-matched subsamples for the three redshift bins, from which one can already observe that the control sample omits a large fraction of the lower-mass spirals, particularly in the intermediate and high redshift bins.
A Kolmogorov-Smirnov (K-S) test supports that these pairs of stellar mass distributions have been drawn from the same parent sample, with $p$-values of $P_{\rm KS}$= 0.69, 0.81, and 0.24 respectively.

The immediate step following the successful construction of the mass-matched control sample would be to compute its bar fraction, The random nature of the selection process, however, implies that the measured bar fraction will fluctuate for different randomly selected samples.
For this reason, we obtained an accurate representation of the bar fraction for the mass-matched control sample as follows:
(1) we randomly select a control sample of spirals meeting the criteria described above, and compute its bar fraction;
(2) we repeat this process one thousand times;
and (3) we calculate the mean bar fraction among all control samples for each redshift bin.
The standard error in this particular $f_{bar}$ will be given by the standard deviation among the one thousand measured bar fractions.

\begin{figure}[t*]
\begin{center}
\includegraphics[scale=0.65]{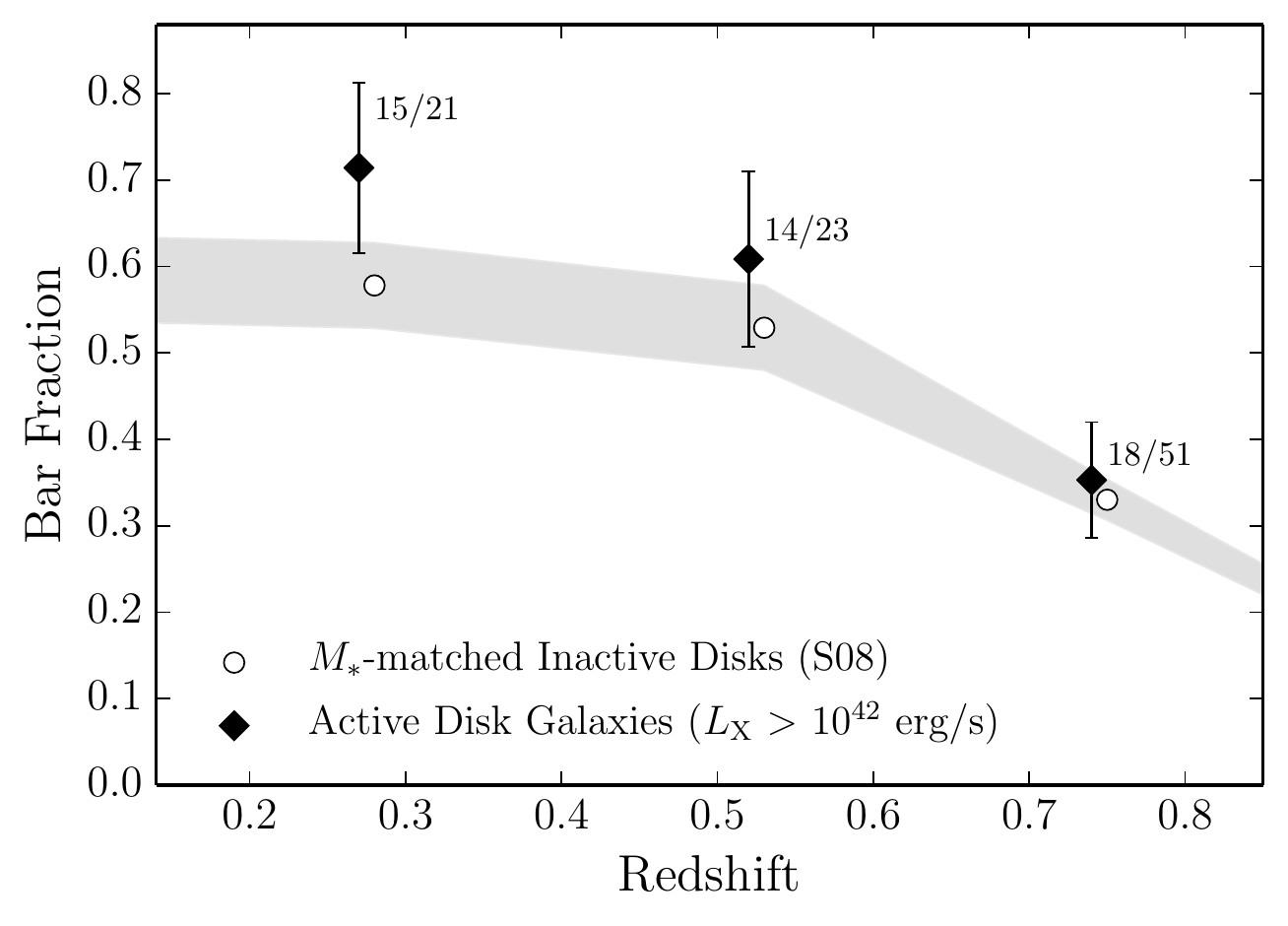}
\caption{
As Figure \ref{fig_bf}, but now comparing the bar fraction of active galaxies to that of inactive galaxies matched in stellar mass.
The mass-matched bar fraction corresponds to the mean of one thousand control samples drawn from the parent sample of inactive disk galaxies from \citetalias{sheth08}.
Each control sample was built by randomly selecting 3 galaxies per AGN, chosen to match in redshift and stellar mass.
\label{fig_bfcs}}
\end{center}
\end{figure}

The mean bar fraction of the mass-matched control samples is presented in Table \ref{tbl_1} and in Figure \ref{fig_bfcs}, where we show it along with the AGN bar fraction.
The resulting $f_{\rm bar}$ appears ``flatter" across the redshift range probed, in agreement with the weaker evolution of the bar fraction for the most massive galaxies.
As expected, at the low redshift bin, the bar fraction of inactive disks remains almost unchanged, yet as one moves toward the higher redshift bins, the bar fraction of the $M_{\ast}$-matched sample is enhanced with respect to the global \citetalias{sheth08} sample from Figure \ref{fig_bf}, and the difference with the AGN bar fraction is largely suppressed.
Considering that the tentative yet not statistically significant enhancement in the bar fraction remains for the lowest redshift bin, with a difference between bar fractions of $\Delta f_{\rm bar} = 0.13 \pm 0.11$ ($\sim$1.2$\sigma$), we choose to perform a more thorough analysis on our results, and provide precise constraints as far as our data allows on the true value of $\Delta f_{\rm bar}$.

\subsubsection{The Difference Between Bar Fractions}

\begin{figure}[t*]
\begin{center}
\includegraphics[scale=0.65]{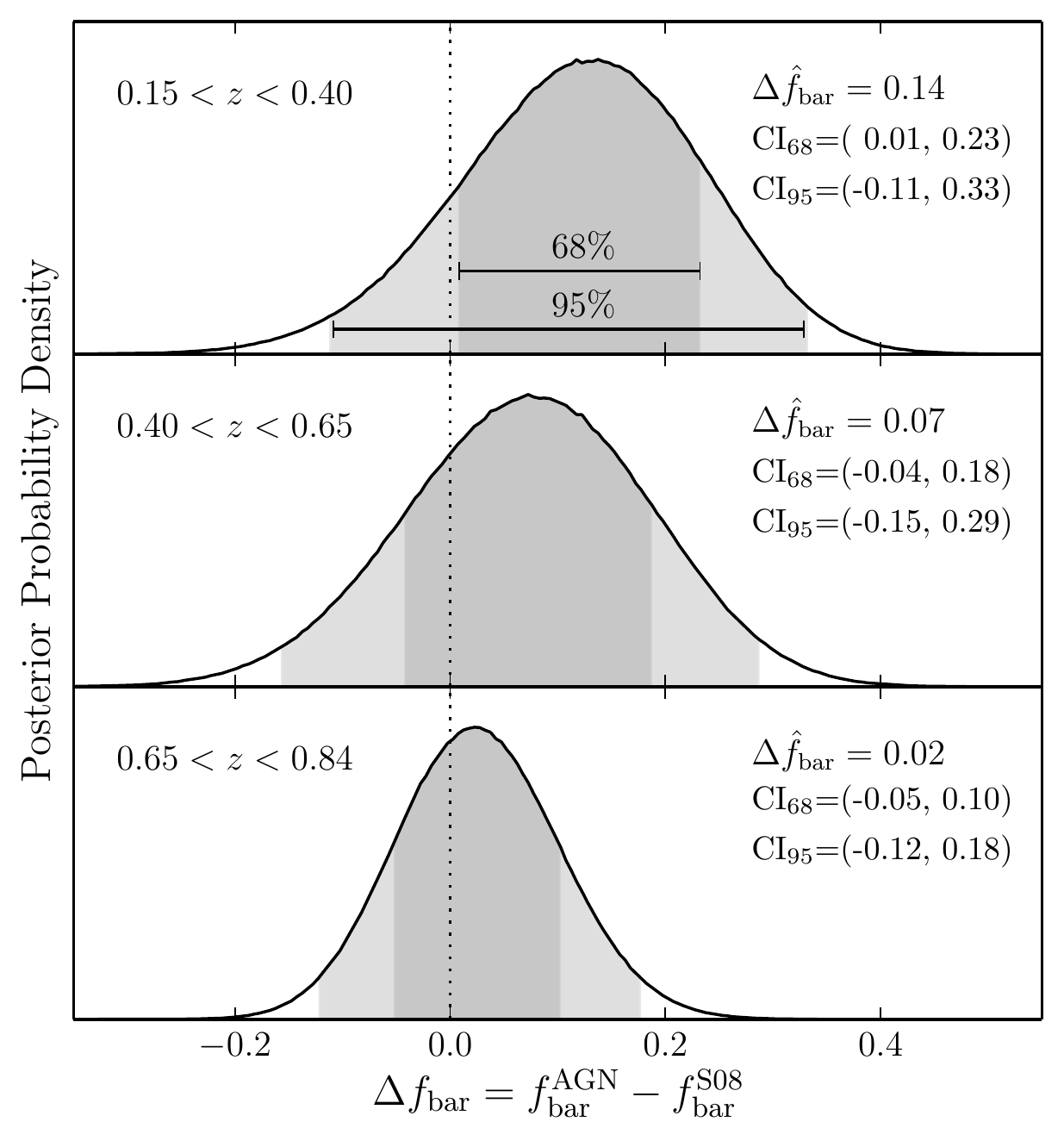}
\caption{
Posterior probability distributions of the difference between the bar fractions of the AGN and the mass-matched control sample, $\Delta f_{\rm bar}$, for the three redshift bins.
In each panel we include maximum-likelihood estimator between bar fractions, $\Delta\hat{f}_{\rm bar}$, and the 68\% and 95\% credible intervals of $\Delta f_{\rm bar}$, CI$_{68}$ and CI$_{95}$ respectively, also illustrated by the shaded areas.
\label{fig_probs}}
\end{center}
\end{figure}

Below, we provide quantitative constraints on the actual difference between the bar fractions of the AGN and mass-matched \citetalias{sheth08} samples, $\Delta f_{bar}$.
For any given sample of galaxies, we can construct a Bayesian model for the posterior probability distribution of its ``true'' bar fraction based on the observed data ($N_{\rm bar}$ and $N_{\rm tot}$ from Table~\ref{tbl_1})\footnote{See e.g., \citet{andrae10} and \citet{cameron11} for further details on this methodology, and \citet{cisternas11a} for an example with merger fractions.}
:
\begin{equation*}
P(f_{\rm bar}|N_{\rm bar}, N_{\rm tot}) \propto P(N_{\rm bar}, N_{\rm tot}|f_{\rm bar}) \times P(f_{\rm bar})
\end{equation*}
where the likelihood $P(N_{\rm bar}, N_{\rm tot}|f_{\rm bar})$ corresponds to a binomial probability distribution which, after normalization, takes the form of a beta distribution, $Beta(\alpha,\beta)$, with parameters $\alpha=N_{\rm bar}+1$, and $\beta=N_{\rm tot}-N_{\rm bar}+1$.
Assuming a uniform prior, i.e., $P(f_{\rm bar}) = Beta(1,1)$, the above likelihood will be equivalent to the posterior probability of $f_{\rm bar}$.
Knowing how to obtain the probability distribution of each bar fraction, we construct the posterior probability of $\Delta f_{\rm bar}$ for the three redshift bins through a Monte Carlo simulation as follows: 
(1) separately, we sample the bar fraction of the AGN and the mass-matched control sample from their corresponding beta distributions,
(2) we compute the difference, $\Delta f_{\rm bar}$, between the above bar fractions,
and (3) we repeat this process five million times.
The resulting posterior probability distributions of the difference between bar fractions are shown in Figure~\ref{fig_probs}.
For each distribution, we report the maximum-likelihood estimator for the difference between bar fractions, $\Delta \hat{f}_{\rm bar}$, and the 68\% ($\approx$$1\sigma$) and 95\% ($\approx$$2\sigma$) credible intervals.

For the lowest redshift bin, which showed the most intriguing difference between bar fractions,
we can see from our results that the 68\% credible interval excludes the $\Delta f_{\rm bar}=0$ line, favoring the possibility that the AGN bar fraction is higher than that of the mass-matched inactive galaxies.
However, using a more standard and conservative credible interval of 95\%, the difference between bar fractions {\em always} includes $\Delta f_{\rm bar}=0$, as it does for the higher-redshift bins, and therefore we cannot rule out the possibility that the bar fractions of AGN and inactive galaxies are equivalent with our current data set.

An independent look at the AGN bar fraction at $z>0$ was recently carried out by \citet{cheung15}, who studied the fraction of barred active galaxies at $0.2<z<1.0$ using classifications from Galaxy Zoo Hubble project in the COSMOS field, as well as from the EGS and GOODS-S fields.
The bar fraction of their sample of 119 AGNs shows no overall enhancement with respect to matched inactive galaxies, in broad agreement with our findings.

\subsection{Bar Presence and Nuclear Activity}

The above results show that, at face value, active disk galaxies are barred more often than inactive ones.
The observed excess of bars, however, can be largely explained by the fact that AGNs typically reside in massive galaxies, which in turn have a higher bar fraction than less massive systems.
With this taken into account, we find no substantial evidence to rule out the possibility that AGN and similarly massive inactive galaxies are barred with comparable frequencies.
Nevertheless, the fact that stellar bars {\em are} able to bring large supplies of gas to the center of galaxies, together with the observed occurrence of AGN activity in barred galaxies, could still suggest that bars may play a role in the nuclear fueling of at least {\em some} galaxies.
In this respect, to further understand whether stellar bars have an impact on the ongoing AGN activity, it is relevant to directly compare the nuclear activity levels in barred and unbarred systems.

\begin{figure}[t]
\begin{center}
\includegraphics[scale=0.65]{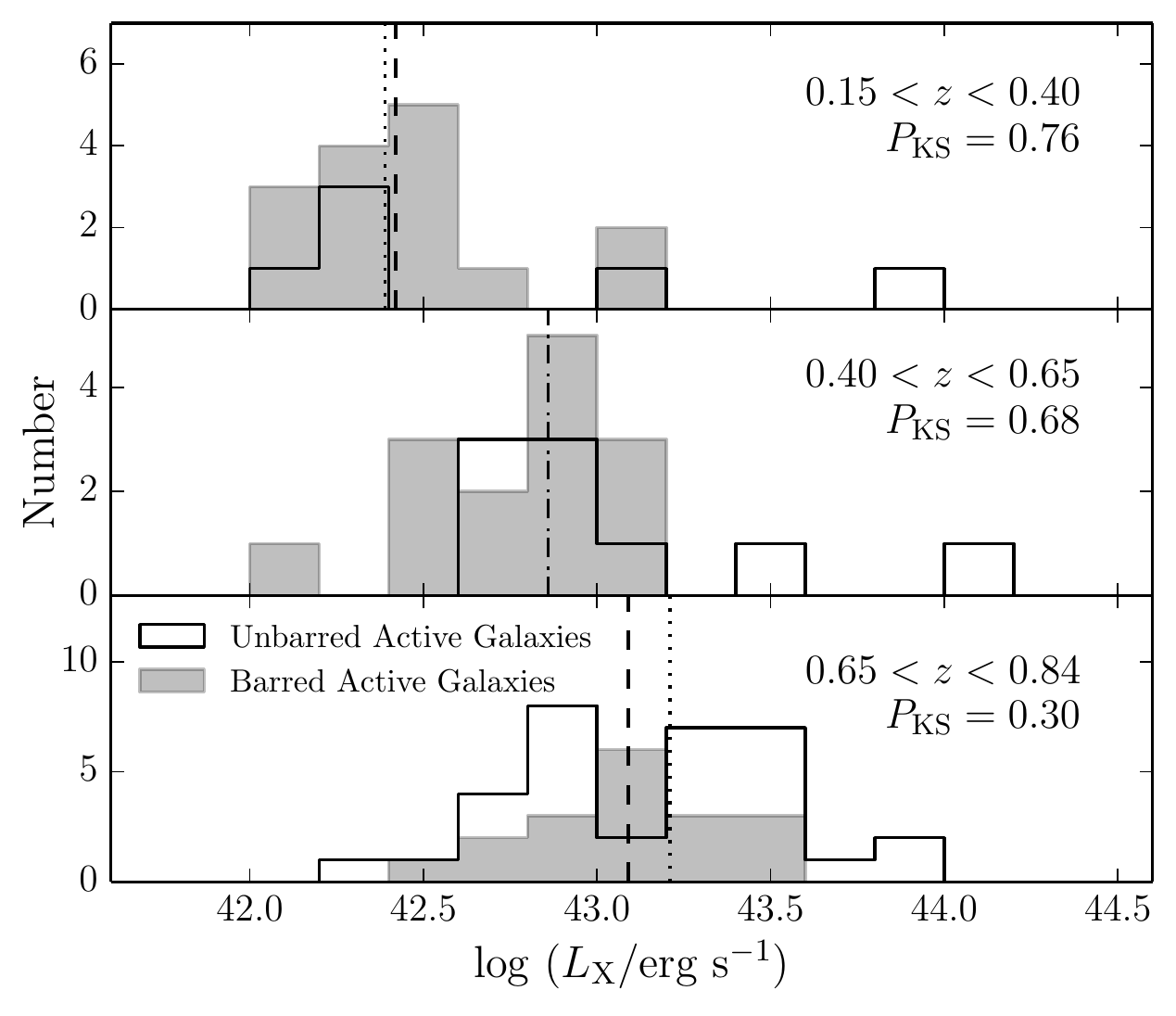}
\caption{X-ray luminosity distributions of barred (gray histograms) and unbarred (empty histograms) active galaxies over three redshift intervals, with their corresponding $p$-value from K-S test, $P_{\rm KS}$. For each distribution, we show its median as vertical dashed (barred) and dotted (unbarred) lines. \label{fig_lxhist}}
\end{center}
\end{figure}

\begin{figure}[t]
\begin{center}
\includegraphics[scale=0.65]{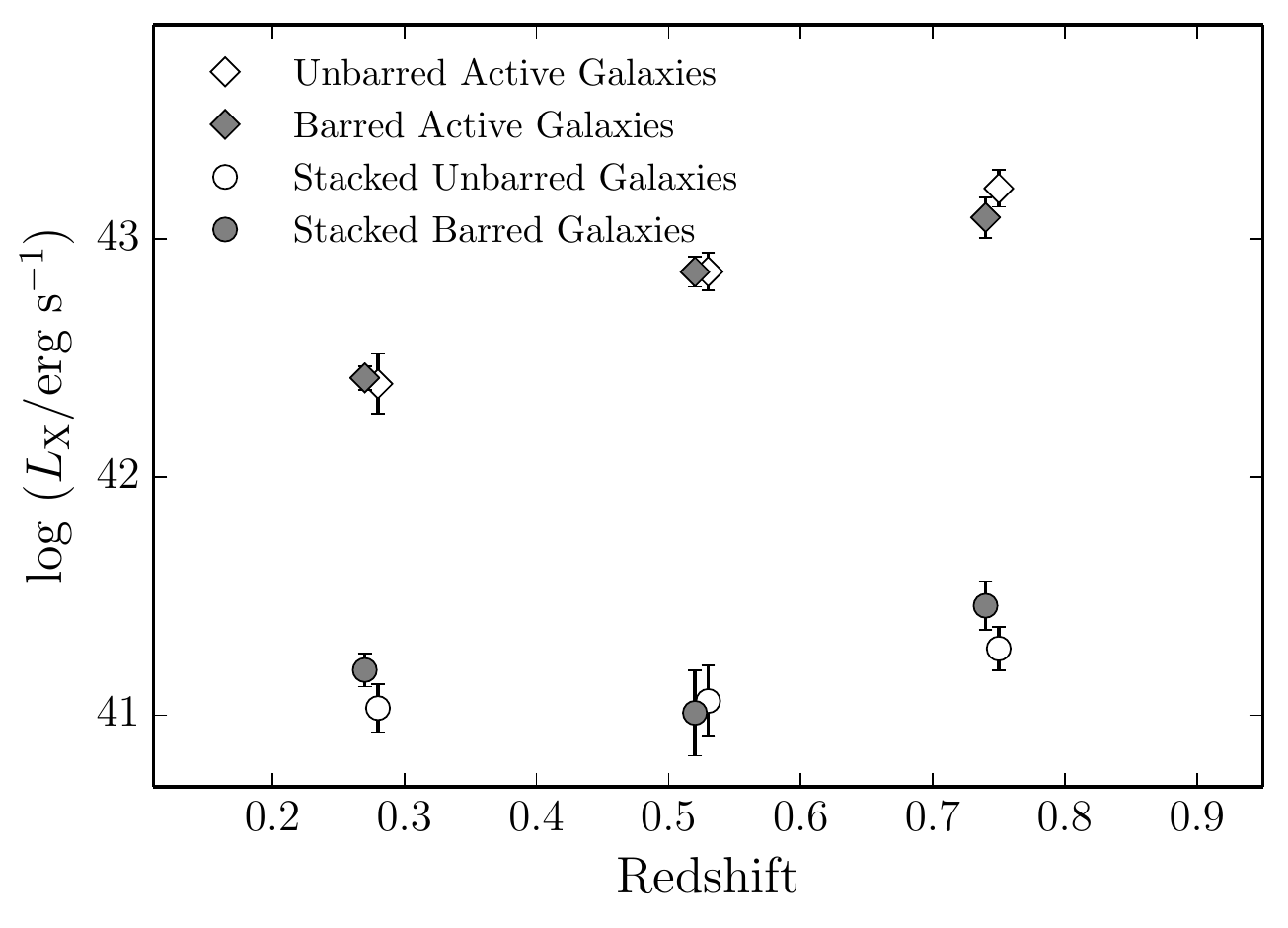}
\caption{Average X-ray luminosities of barred (filled symbols) and unbarred (empty symbols) galaxies in three redshift bins for X-ray-detected active galaxies (diamonds) and undetected stacked galaxies (circles).\label{fig_lxplot}}
\label{default}
\end{center}
\end{figure}

We investigate whether in our sample of active $L^{\ast}$ disk galaxies, AGN activity is enhanced by the presence of a bar.
To this extent, we use the 2--10 keV X-ray luminosities as a proxy of AGN strength which, although it also includes some nominal contribution from star formation, at these luminosity regime ($L_{\rm X} \ge10^{42}$ erg s$^{-1}$) can be safely attributed primarily to BH accretion.
In Figure \ref{fig_lxhist} we show the distributions of X-ray luminosities in three redshift bins.
At a first glance, no major difference can be seen between the distributions of barred (gray histograms) and unbarred (empty histograms) active galaxies.
Quantitatively, the median $L_{\rm X}$ values of barred and unbarred galaxies are statistically indistinguishable in all three redshift bins, as illustrated by the vertical lines in each panel, also presented in Figure \ref{fig_lxplot}.
Performing a K-S test further strengthens this point, by showing a high probability that each pair of luminosity distributions is drawn from the same parent distribution ($P_{\rm KS}$ = 0.76, 0.68, and 0.30).
These results suggest that the presence of a large scale bar has no particular influence on the ongoing BH activity, as also observed in nearby galaxies \citep{ho97b,cisternas13}, yet our present sample is probing AGN luminosities at least an order of magnitude higher than typical local Seyferts.

\subsubsection{X-ray Stacking of Inactive Spirals}

As a second test to trace the influence of bars on nuclear activity, we unveil the hitherto hidden X-ray emission of galaxies without individual {\em Chandra} detections.
We accomplish this by resorting to a stacking analysis of {\em Chandra} observations using the latest version (v4.1) of CSTACK tool\footnote{http://cstack.ucsd.edu}.
From the \citetalias{sheth08} sample of $L^{\ast}$ disk galaxies, we construct subsamples of barred and unbarred galaxies separately, for the three redshift bins, i.e., 0.15--0.40, 0.40--0.65, and 0.65--0.84.
The X-ray stacking procedure yields the average X-ray emission of these individually undetected galaxies, and serves as a way of probing not necessarily luminous BH activity, but rather low-level AGNs and enhanced star formation activity.
Any galaxy already in the X-ray source catalog is excluded from the stacking analysis, as well as those targets located near resolved sources.
We assess the significance of the stacking by comparing our detection to the standard error from 500 resampled stacked count rates from a bootstrap procedure.
The total number of sources per stack is given in Table \ref{tbl_2}, together with the significance of the detection above the noise for the soft and hard bands.

If bar-driven inflows are indeed able to have an impact on the innermost regions of a galaxy, one may expect barred galaxies to show, on average, higher levels of absorption in the X-ray spectrum due to obscuring material.
In this respect, we compare the levels of nuclear obscuration between stacks of barred and unbarred galaxies by computing their X-ray hardness ratios ($HR$), given by $HR=(H-S)/(H+S)$, where $H$ and $S$ are the count rates in the soft (0.5--2 keV) and hard (2--8 keV) {\em Chandra} bands respectively.
While each individual stack shows a decent detection ($>$2.4$\sigma$) in the soft band, only two stacks have significant detections in the hard band.
For those cases without detections, we provide an upper limit on the $HR$ by assuming a count rate in the hard band at the  $2\sigma$ level.
The $HR$ values are presented in Table \ref{tbl_2}.

\begin{deluxetable}{lcccr@{}lc}
\tabletypesize{\scriptsize}
\tablecaption{X-ray Stacking Results\label{tbl_2}}
\tablewidth{0.45\textwidth}
\tablehead{
$N_{\rm Stack}$\tablenotemark{a} & $\tilde{z}$\tablenotemark{b} & \multicolumn{2}{c}{Detection ($\sigma$)\tablenotemark{c}} & \multicolumn{2}{c}{$HR$\tablenotemark{d}} & log $L_{\rm X}$\\
 & & [Soft] & [Hard] & & & (erg s$^{-1}$)}
\startdata
Unbarred Galaxies\\
81  &  0.35 & 4.4 & 1.5 & $<$-&0.16 & $41.0\pm0.1$\\
186  &  0.55 & 2.8 & 2.7 & & 0.19 & $41.1\pm0.2$ \\
362 &  0.71 & 5.1 & $<$1 & $<$-&0.16 & $41.3\pm0.1$\\ \hline
Barred Galaxies\\
127  &  0.34 & 6.3 & 1.9 & $<$-&0.40 & $41.2\pm0.1$ \\
148  &  0.52 & 2.4 & $<$1 & $<$\,&0.13 &$41.0\pm0.2$\\
113 &  0.70 & 4.2 & 2.7 & & 0.08 & $41.5\pm0.1$
\enddata

\tablenotetext{a}{Number of accepted galaxies in stack.}
\tablenotetext{b}{Median redshift of galaxies in stack, used to compute fluxes and luminosities.}
\tablenotetext{c}{Detection significance above the noise for each band.}
\tablenotetext{d}{Hardness ratio, defined as $(H-S)/(H+S)$, where $H$ and $S$ are the count rates in the hard and soft bands respectively.}
\end{deluxetable}

Because of the lack of consistent hard band detections, we can only make crude assumptions on the hardness of the spectra.
Unobscured AGNs tend to have $HR$ values below -0.2, corresponding to soft, unabsorbed spectra \citep{hasinger08}.
The $HR$ values of our stacks are not particularly hard, with values always below 0.2, even for the stacks with the clearest detections in the hard band.
Comparatively, no significant differences are observed between barred and unbarred subsamples, suggesting that the nuclear conditions of these galaxies are not intrinsically distinct.

To investigate the levels of nuclear activity, we compare the average X-ray luminosities between stacks.
Because of the scant hard band detections, we use the soft band count rates to estimate 2--10 keV fluxes and luminosities.
We consider Galactic absorption, and assume an absorbed power-law model with spectral index $\Gamma=1.4$ and intrinsic column density $N_{\rm H}=10^{22}\,{\rm cm}^{-2}$, a representative value for moderate-luminosity AGNs \citep[e.g.,][]{bauer04}.
The estimated luminosities are presented in Table \ref{tbl_2}, and shown in Figure \ref{fig_lxplot}.
As it can be seen from our results, we find no significant difference between the $L_{\rm X}$ values of the barred and unbarred stacks that could hint at obscured BH growth or enhanced star formation going on in barred galaxies.

\subsection{The Limited Role of Bars}

Our results show that
for our sample of moderate luminosity X-ray-selected AGNs ($L_{\rm X}\sim10^{43}$ erg s$^{-1}$) hosted in luminous disk galaxies at $0.15<z<0.84$,
(1) 
the fraction of barred active galaxies evolves in a similar fashion as that of inactive galaxies,
(2)
active and inactive galaxies are barred with comparable frequencies,
and (3) barred and unbarred galaxies show indistinguishable levels of AGN activity.

While we showed that we cannot rule out that the fractions of active and inactive barred galaxies are the same at all redshifts, it is still interesting to compare the difference we found at the lowest redshift bin, $\Delta \hat{f}_{\rm bar}=0.13\pm0.11$, with other studies from the local universe.
\citet{knapen00} found bar fractions for their samples of nearby Seyfert galaxies and control inactive galaxies of $0.79\pm0.08$, and $0.59\pm0.09$ respectively, which translates into a difference of $0.20\pm0.12$, or 1.7$\sigma$.
Following up on this intriguing result, \citet{laine02} increased the sample size, and obtained bar fractions of $0.73\pm0.06$ and $0.50\pm0.07$ for Seyfert and inactive galaxies respectively, corresponding to a difference of $0.23\pm0.09$ and yielding a higher significance of 2.5$\sigma$.
We caution that direct comparisons between studies are not straightforward given the different characteristics between samples.
Nonetheless, it is still interesting as it hints at a trend of an increasing difference between bar fractions with decreasing redshift, yet to be confirmed with larger samples.

The results from nearby Seyfert galaxies could also provide clues on why there is no enhancement in the AGN bar fraction at high redshift, but a difference seems to appear as one approaches $z=0$:
the average AGN luminosities probed in our sample increase with redshift (see Figure \ref{fig_lxplot}), and therefore it is possible that  bar-driven inflows favor the occurrence of AGNs up to a certain luminosity.
For the high-luminosity end, the fueling requirements might be intrinsically different, and other mechanisms might be more efficient at providing the necessary supply of gas without the need of bars (see next Section).
This could account for the mild, yet not significant, enhancement of the AGN bar fractions in our low redshift subsample as well as in local Seyferts (i.e., the low-luminosity end), and explain why the enhancement is washed out toward higher redshifts and luminosities.

Based on what we know from the literature, one would expect stellar bars to increase the odds of AGN activity occurring:
bar-driven gas inflows from kpc scales result in the buildup of central concentrations of gas \citep{sakamoto99, sheth05}, star formation \citep{martin95,ho97b}, and subsequent pseudobulge growth \citep{kormendy&kennicutt04}.
Such conditions are expected to increase the likelihood of, e.g., the formation of a gaseous disk within the central hundred pc, which could be subject to further instabilities leading to inflows that would fuel the central BH \citep{shlosman89}.
The uncertainties inherent to our sample size, however, do not allow us to confirm a difference between the bar fractions of AGN and inactive galaxies.
Furthermore, the indistinguishable levels of AGN activity observed in barred and unbarred galaxies, suggests that the inflow of gas from kpc scales is not a necessary condition for BH fueling, and that at $<$100 pc scales the transport mechanisms tend to be disconnected from the large-scale structure of the galaxy.
This idea was shown by \citet{martini03b} through their study of nuclear spiral structure---a possible mechanism to transport material inward near the vicinity of a central BH.
They found that most active galaxies feature nuclear dust spiral structure at their centers, yet it tends to be rather asymmetric and not connected to a larger stellar bar, suggesting a possible in-situ formation scenario for these nuclear spirals.

Another aspect worth mentioning is that, even if there was a direct link between bars and AGN,
detecting correlations is particularly difficult even at high redshift.
This directly relates to the fact that an AGN is a rather short-lived, episodic phenomenon, with typical lifetimes of 10--100 Myr \citep[e.g.,][]{martini04_lifetime} which is significantly shorter than periods of enhanced star formation.
This implies that it is not strange to find correlations regarding star formation in barred galaxies, yet AGN activity is far more elusive, making any observable link much more subtle.
Additionally, as we discuss below, AGN activity is also observed in unbarred galaxies, meaning that other mechanisms can also provide the necessary gas supply, and further dilute any connection between bars and AGNs. 

\subsection{Most Active Disk Galaxies are Unbarred}

Looking at our results from a different perspective, it is important to mention that roughly 60\% of disk-hosted AGNs at $z>0.4$ appear to lack large-scale bars, with this number being a lower limit if one extrapolates the observed trend out to higher redshifts, beyond $z=0.84$.
Therefore, even if there was a causal connection between bars and AGN activity for at least some galaxies, other alternatives to bring enough gas down to the nuclear regions should be investigated to account for the fueling of these AGNs in unbarred hosts.

The interpretation of these findings requires an understanding of the nature of stellar bars: they are a reliable indicator of a ``mature" i.e., dynamically cold, stable disk.
In other words, bars do not form in ``hot" disks, i.e., rather turbulent ones, with a dominant velocity dispersion component.
Therefore, this could be directly connected with the decrease in the bar fraction at higher redshifts and hot disks being more common there \citep{sheth12}.
In fact, 
due to the higher gas fractions at earlier cosmic times \citep{tacconi10,daddi10a},
star-forming disk galaxies at $z\sim2$ tend to be quite turbulent, gas-rich, and unstable, which can lead to fragmentation and formation of dense clouds and giant clumps \citep{genzel08, bournaud11}.
At $z<1$, the gas fractions drop and disks tend to become more stable and less clumpy.
As the fraction of clumpy galaxies decreases from $\sim$35\% at $z=1$ down to $\sim$5\% at $z=0.2$ \citep{murata14}, the bar fraction increases by a factor of 3.

Could the dynamical conditions of disk galaxies at higher redshifts be also related to the increasing fraction of AGN activity in unbarred galaxies?
Massive clouds and clumpy structures in a disk can migrate radially by themselves, or induce angular momentum transfer outward, resulting in gas inflows which could promote bulge, and potentially BH, growth \citep{dekel09,bournaud11,gabor&bournaud13}.
A link between clumpy disks and AGN activity at $z\sim0.7$ has been reported by \citet{bournaud12}, yet for luminosities well below our cutoff, i.e., $L_{\rm X}<10^{42}$ erg s$^{-1}$, arguing that this mechanism would only account for very modest levels of activity at these redshifts.
At $z\sim2$, however, \citet{trump14} found similar levels of AGN activity in clumpy and smooth disks, ruling out violent disk instabilities as a dominant or more efficient BH fueling mechanism. 

While a more in depth analysis is beyond the scope of this Paper, our sample only shows a minority of galaxies with clear clumps.
The transient nature of these features, as well as the limited AGN lifetime, makes it difficult to find a direct connection between AGN activity and clumpy structure.
The AGN luminosities probed here, however, may not need of massive clumps to be accounted for, and the stochastic accretion of a few giant molecular clouds could still provide the necessary fuel \citep[][]{hopkins06_stochastic}.
Clump-free, smooth disks are not necessarily completely dynamically-cold, and the existence of a turbulent component could favor the occurrence of these random events.
A promising next step would be a comparative study on the kinematics and gas content of barred and unbarred active galaxies at high redshift.
This could provide valuable clues on whether the dynamical state of the disk is indeed a fundamental factor which provides the suitable conditions for AGN fueling on the majority of active galaxies.

\section{Conclusions}

At least over the last seven billion years, a significant fraction of the BH growth is occurring in seemingly undisturbed disk galaxies.
Stellar bars, highly common in the local universe, are often invoked as a mechanism which can deliver a fresh supply gas to feed these AGNs, yet no observational studies have so far probed their actual relevance beyond $z=0$.

In this Paper, we have carefully constructed a sample of 95 AGNs,
selected based on their 2--10 keV X-ray luminosities ($L_{\rm X}\ge10^{42}$ erg s$^{-1}$),
and hosted in luminous moderately inclined disk galaxies over $0.15<z<0.84$ with the goal of investigating the impact of bars on nuclear activity, as well as the redshift evolution of the bar fraction of active galaxies.
Our results can be summarized as follows:

\begin{itemize}

\item[1.] 
The bar fraction of active galaxies declines with increasing redshift, from 71\%$\pm$10\%  at $z\sim0.3$ to 35\%$\pm$7\% $z\sim0.8$, showing a similar evolution as that of ``normal" spiral galaxies.

\item[2.]
When directly compared against inactive disk galaxies, the fraction of bars is higher in active disk galaxies.
Most of this enhancement, however, is because AGNs in our sample are typically hosted by rather massive galaxies, which in turn have a higher bar fraction.
When compared against a mass-matched sample of inactive galaxies, the enhancement is largely suppressed, and we cannot rule out that the bar fractions of active and inactive galaxies are equivalent.

\item[3.]
The strength of the AGN activity is not influenced by the presence of a bar.
Barred and unbarred active galaxies show equivalent $L_{\rm X}$ distributions over the redshift range probed.
Through an X-ray stacking analysis of inactive barred and unbarred galaxies, we find comparable levels of X-ray emission, likely attributable to star formation.

\item[4.]
From our findings above, we conclude that the occurrence and the efficiency of AGN activity is independent of the large-scale structure of the galaxy.
The role of bars may be confined to bringing large supplies of gas to the central regions which could eventually be transported further down by other more relevant mechanisms at work at the $<$100 pc scale.

\item[5.] 
Viewing our results from a different angle, we find that an increasing fraction of unbarred active galaxies at higher redshifts, which we interpret to be related to the rather unstable dynamical conditions of disk galaxies at earlier cosmic times.

\end{itemize}


\acknowledgments
M. C. thanks Takamitsu Miyaji for helping with CSTACK, Anton Koekemoer for useful comments, and the anonymous referee for a constructive report.
We gratefully acknowledge the COSMOS bars team for providing the original classifications, and the contributions of the entire COSMOS collaboration consisting of more than 100 scientists.
More information about the COSMOS survey is available at http://cosmos.astro.caltech.edu.
J. H. K. acknowledges financial support to the DAGAL network from the People 
Programme (Marie Curie Actions) of the European UnionÕs Seventh 
Framework Programme FP7/2007-2013/ under REA grant agreement number 
PITN-GA-2011-289313, and from the Spanish MINECO under grant number AYA2013-41243-P.
F. C. acknowledges financial support by the NASA grant GO3-14150C.
The National Radio Astronomy Observatory is a facility of the National Science Foundation operated under cooperative agreement by Associated Universities, Inc.

{\it Facilities:}  \facility{CXO}, \facility{{\it HST} (ACS)}






\begin{thebibliography}{}
\expandafter\ifx\csname natexlab\endcsname\relax\def\natexlab#1{#1}\fi

\bibitem[{{Andrae}(2010)}]{andrae10}
{Andrae}, R. 2010, ArXiv:1009.2755, arXiv:1009.2755

\bibitem[{{Athanassoula}(1992)}]{athanassoula92b}
{Athanassoula}, E. 1992, \mnras, 259, 345

\bibitem[{{Barnes} \& {Hernquist}(1991)}]{barnes&hernquist91}
{Barnes}, J.~E., \& {Hernquist}, L.~E. 1991, \apjl, 370, L65

\bibitem[{{Bauer} {et~al.}(2004){Bauer}, {Alexander}, {Brandt}, {Schneider},
  {Treister}, {Hornschemeier}, \& {Garmire}}]{bauer04}
{Bauer}, F.~E., {Alexander}, D.~M., {Brandt}, W.~N., {et~al.} 2004, \aj, 128,
  2048

\bibitem[{{Bournaud} {et~al.}(2011){Bournaud}, {Dekel}, {Teyssier}, {Cacciato},
  {Daddi}, {Juneau}, \& {Shankar}}]{bournaud11}
{Bournaud}, F., {Dekel}, A., {Teyssier}, R., {et~al.} 2011, \apjl, 741, L33

\bibitem[{{Bournaud} {et~al.}(2012){Bournaud}, {Juneau}, {Le Floc'h},
  {Mullaney}, {Daddi}, {Dekel}, {Duc}, {Elbaz}, {Salmi}, \&
  {Dickinson}}]{bournaud12}
{Bournaud}, F., {Juneau}, S., {Le Floc'h}, E., {et~al.} 2012, \apj, 757, 81

\bibitem[{{Cameron}(2011)}]{cameron11}
{Cameron}, E. 2011, PASA, 28, 128

\bibitem[{{Cameron} {et~al.}(2010){Cameron}, {Carollo}, {Oesch}, {Aller},
  {Bschorr}, {Cerulo}, {Aussel}, {Capak}, {Le Floc'h}, {Ilbert}, {Kneib},
  {Koekemoer}, {Leauthaud}, {Lilly}, {Massey}, {McCracken}, {Rhodes},
  {Salvato}, {Sanders}, {Scoville}, {Sheth}, {Taniguchi}, \&
  {Thompson}}]{cameron10}
{Cameron}, E., {Carollo}, C.~M., {Oesch}, P., {et~al.} 2010, \mnras, 409, 346

\bibitem[{{Canalizo} \& {Stockton}(2001)}]{canalizo&stockton01}
{Canalizo}, G., \& {Stockton}, A. 2001, \apj, 555, 719

\bibitem[{{Cheung} {et~al.}(2015){Cheung}, {Trump}, {Athanassoula}, {Bamford},
  {Bell}, {Bosma}, {Cardamone}, {Casteels}, {Faber}, {Fang}, {Fortson},
  {Kocevski}, {Koo}, {Laine}, {Lintott}, {Masters}, {Melvin}, {Nichol},
  {Schawinski}, {Simmons}, {Smethurst}, \& {Willett}}]{cheung15}
{Cheung}, E., {Trump}, J.~R., {Athanassoula}, E., {et~al.} 2015, \mnras, 447,
  506

\bibitem[{{Cisternas} {et~al.}(2011){Cisternas}, {Jahnke}, {Inskip},
  {Kartaltepe}, {Koekemoer}, {Lisker}, {Robaina}, {Scodeggio}, {Sheth},
  {Trump}, {Andrae}, {Miyaji}, {Lusso}, {Brusa}, {Capak}, {Cappelluti},
  {Civano}, {Ilbert}, {Impey}, {Leauthaud}, {Lilly}, {Salvato}, {Scoville}, \&
  {Taniguchi}}]{cisternas11a}
{Cisternas}, M., {Jahnke}, K., {Inskip}, K.~J., {et~al.} 2011, \apj, 726, 57

\bibitem[{{Cisternas} {et~al.}(2013){Cisternas}, {Gadotti}, {Knapen}, {Kim},
  {D{\'{\i}}az-Garc{\'{\i}}a}, {Laurikainen}, {Salo},
  {Gonz{\'a}lez-Mart{\'{\i}}n}, {Ho}, {Elmegreen}, {Zaritsky}, {Sheth},
  {Athanassoula}, {Bosma}, {Comer{\'o}n}, {Erroz-Ferrer}, {Gil de Paz}, {Hinz},
  {Holwerda}, {Laine}, {Meidt}, {Men{\'e}ndez-Delmestre}, {Mizusawa},
  {Mu{\~n}oz-Mateos}, {Regan}, \& {Seibert}}]{cisternas13}
{Cisternas}, M., {Gadotti}, D.~A., {Knapen}, J.~H., {et~al.} 2013, \apj, 776,
  50

\bibitem[{{Civano} {et~al.}(2012){Civano}, {Elvis}, {Brusa}, {Comastri},
  {Salvato}, {Zamorani}, {Aldcroft}, {Bongiorno}, {Capak}, {Cappelluti},
  {Cisternas}, {Fiore}, {Fruscione}, {Hao}, {Kartaltepe}, {Koekemoer}, {Gilli},
  {Impey}, {Lanzuisi}, {Lusso}, {Mainieri}, {Miyaji}, {Lilly}, {Masters},
  {Puccetti}, {Schawinski}, {Scoville}, {Silverman}, {Trump}, {Urry},
  {Vignali}, \& {Wright}}]{cosmos_civano12}
{Civano}, F., {Elvis}, M., {Brusa}, M., {et~al.} 2012, \apjs, 201, 30

\bibitem[{{Coelho} \& {Gadotti}(2011)}]{coelho11}
{Coelho}, P., \& {Gadotti}, D.~A. 2011, \apjl, 743, L13

\bibitem[{{Daddi} {et~al.}(2010){Daddi}, {Bournaud}, {Walter}, {Dannerbauer},
  {Carilli}, {Dickinson}, {Elbaz}, {Morrison}, {Riechers}, {Onodera}, {Salmi},
  {Krips}, \& {Stern}}]{daddi10a}
{Daddi}, E., {Bournaud}, F., {Walter}, F., {et~al.} 2010, \apj, 713, 686

\bibitem[{{de Vaucouleurs}(1963)}]{devaucouleurs63}
{de Vaucouleurs}, G. 1963, \apjs, 8, 31

\bibitem[{{Dekel} {et~al.}(2009){Dekel}, {Sari}, \& {Ceverino}}]{dekel09}
{Dekel}, A., {Sari}, R., \& {Ceverino}, D. 2009, \apj, 703, 785

\bibitem[{{Draper} \& {Ballantyne}(2012)}]{draper12}
{Draper}, A.~R., \& {Ballantyne}, D.~R. 2012, \apj, 751, 72

\bibitem[{{Ellison} {et~al.}(2011){Ellison}, {Nair}, {Patton}, {Scudder},
  {Mendel}, \& {Simard}}]{ellison11}
{Ellison}, S.~L., {Nair}, P., {Patton}, D.~R., {et~al.} 2011, \mnras, 416, 2182

\bibitem[{{Elvis} {et~al.}(2009){Elvis}, {Civano}, {Vignali}, {Puccetti},
  {Fiore}, {Cappelluti}, {Aldcroft}, {Fruscione}, {Zamorani}, {Comastri},
  {Brusa}, {Gilli}, {Miyaji}, {Damiani}, {Koekemoer}, {Finoguenov}, {Brunner},
  {Urry}, {Silverman}, {Mainieri}, {Hasinger}, {Griffiths}, {Carollo}, {Hao},
  {Guzzo}, {Blain}, {Calzetti}, {Carilli}, {Capak}, {Ettori}, {Fabbiano},
  {Impey}, {Lilly}, {Mobasher}, {Rich}, {Salvato}, {Sanders}, {Schinnerer},
  {Scoville}, {Shopbell}, {Taylor}, {Taniguchi}, \&
  {Volonteri}}]{cosmos_chandra1}
{Elvis}, M., {Civano}, F., {Vignali}, C., {et~al.} 2009, \apjs, 184, 158

\bibitem[{{Eskridge} {et~al.}(2000){Eskridge}, {Frogel}, {Pogge}, {Quillen},
  {Davies}, {DePoy}, {Houdashelt}, {Kuchinski}, {Ram{\'{\i}}rez}, {Sellgren},
  {Terndrup}, \& {Tiede}}]{eskridge00}
{Eskridge}, P.~B., {Frogel}, J.~A., {Pogge}, R.~W., {et~al.} 2000, \aj, 119,
  536

\bibitem[{{Gabor} \& {Bournaud}(2013)}]{gabor&bournaud13}
{Gabor}, J.~M., \& {Bournaud}, F. 2013, \mnras, 434, 606

\bibitem[{{Gabor} {et~al.}(2009){Gabor}, {Impey}, {Jahnke}, {Simmons}, {Trump},
  {Koekemoer}, {Brusa}, {Cappelluti}, {Schinnerer}, {Smol{\v c}i{\'c}},
  {Salvato}, {Rhodes}, {Mobasher}, {Capak}, {Massey}, {Leauthaud}, \&
  {Scoville}}]{gabor09}
{Gabor}, J.~M., {Impey}, C.~D., {Jahnke}, K., {et~al.} 2009, \apj, 691, 705

\bibitem[{{Garc{\'{\i}}a-Burillo} {et~al.}(2005){Garc{\'{\i}}a-Burillo},
  {Combes}, {Schinnerer}, {Boone}, \& {Hunt}}]{garcia-burillo05}
{Garc{\'{\i}}a-Burillo}, S., {Combes}, F., {Schinnerer}, E., {Boone}, F., \&
  {Hunt}, L.~K. 2005, \aap, 441, 1011

\bibitem[{{Genzel} {et~al.}(2008){Genzel}, {Burkert}, {Bouch{\'e}}, {Cresci},
  {F{\"o}rster Schreiber}, {Shapley}, {Shapiro}, {Tacconi}, {Buschkamp},
  {Cimatti}, {Daddi}, {Davies}, {Eisenhauer}, {Erb}, {Genel}, {Gerhard},
  {Hicks}, {Lutz}, {Naab}, {Ott}, {Rabien}, {Renzini}, {Steidel}, {Sternberg},
  \& {Lilly}}]{genzel08}
{Genzel}, R., {Burkert}, A., {Bouch{\'e}}, N., {et~al.} 2008, \apj, 687, 59

\bibitem[{{Georgakakis} {et~al.}(2009){Georgakakis}, {Coil}, {Laird},
  {Griffith}, {Nandra}, {Lotz}, {Pierce}, {Cooper}, {Newman}, \&
  {Koekemoer}}]{georgakakis09}
{Georgakakis}, A., {Coil}, A.~L., {Laird}, E.~S., {et~al.} 2009, \mnras, 397,
  623

\bibitem[{{Goulding} {et~al.}(2014){Goulding}, {Forman}, {Hickox}, {Jones},
  {Murray}, {Paggi}, {Ashby}, {Coil}, {Cooper}, {Huang}, {Kraft}, {Newman},
  {Weiner}, \& {Willner}}]{goulding14}
{Goulding}, A.~D., {Forman}, W.~R., {Hickox}, R.~C., {et~al.} 2014, \apj, 783,
  40

\bibitem[{{Haan} {et~al.}(2009){Haan}, {Schinnerer}, {Emsellem},
  {Garc{\'{\i}}a-Burillo}, {Combes}, {Mundell}, \& {Rix}}]{haan09}
{Haan}, S., {Schinnerer}, E., {Emsellem}, E., {et~al.} 2009, \apj, 692, 1623

\bibitem[{{Hasinger}(2008)}]{hasinger08}
{Hasinger}, G. 2008, \aap, 490, 905

\bibitem[{{Ho} {et~al.}(1997){Ho}, {Filippenko}, \& {Sargent}}]{ho97b}
{Ho}, L.~C., {Filippenko}, A.~V., \& {Sargent}, W.~L.~W. 1997, \apj, 487, 591

\bibitem[{{Hopkins} \& {Hernquist}(2006)}]{hopkins06_stochastic}
{Hopkins}, P.~F., \& {Hernquist}, L. 2006, \apjs, 166, 1

\bibitem[{{Hopkins} {et~al.}(2014){Hopkins}, {Kocevski}, \&
  {Bundy}}]{hopkins&kocevski14}
{Hopkins}, P.~F., {Kocevski}, D.~D., \& {Bundy}, K. 2014, \mnras, 445, 823

\bibitem[{{Hunt} \& {Malkan}(2004)}]{hunt&malkan04}
{Hunt}, L.~K., \& {Malkan}, M.~A. 2004, \apj, 616, 707

\bibitem[{{Ilbert} {et~al.}(2005){Ilbert}, {Tresse}, {Zucca}, {Bardelli},
  {Arnouts}, {Zamorani}, {Pozzetti}, {Bottini}, {Garilli}, {Le Brun}, {Le
  F{\`e}vre}, {Maccagni}, {Picat}, {Scaramella}, {Scodeggio}, {Vettolani},
  {Zanichelli}, {Adami}, {Arnaboldi}, {Bolzonella}, {Cappi}, {Charlot},
  {Contini}, {Foucaud}, {Franzetti}, {Gavignaud}, {Guzzo}, {Iovino},
  {McCracken}, {Marano}, {Marinoni}, {Mathez}, {Mazure}, {Meneux}, {Merighi},
  {Paltani}, {Pello}, {Pollo}, {Radovich}, {Bondi}, {Bongiorno}, {Busarello},
  {Ciliegi}, {Lamareille}, {Mellier}, {Merluzzi}, {Ripepi}, \&
  {Rizzo}}]{ilbert05}
{Ilbert}, O., {Tresse}, L., {Zucca}, E., {et~al.} 2005, \aap, 439, 863

\bibitem[{{Ilbert} {et~al.}(2009){Ilbert}, {Capak}, {Salvato}, {Aussel},
  {McCracken}, {Sanders}, {Scoville}, {Kartaltepe}, {Arnouts}, {Floc'h},
  {Mobasher}, {Taniguchi}, {Lamareille}, {Leauthaud}, {Sasaki}, {Thompson},
  {Zamojski}, {Zamorani}, {Bardelli}, {Bolzonella}, {Bongiorno}, {Brusa},
  {Caputi}, {Carollo}, {Contini}, {Cook}, {Coppa}, {Cucciati}, {de la Torre},
  {de Ravel}, {Franzetti}, {Garilli}, {Hasinger}, {Iovino}, {Kampczyk},
  {Kneib}, {Knobel}, {Kovac}, {LeBorgne}, {LeBrun}, {F{\`e}vre}, {Lilly},
  {Looper}, {Maier}, {Mainieri}, {Mellier}, {Mignoli}, {Murayama}, {Pell{\`o}},
  {Peng}, {P{\'e}rez-Montero}, {Renzini}, {Ricciardelli}, {Schiminovich},
  {Scodeggio}, {Shioya}, {Silverman}, {Surace}, {Tanaka}, {Tasca}, {Tresse},
  {Vergani}, \& {Zucca}}]{cosmos_ilbert09}
{Ilbert}, O., {Capak}, P., {Salvato}, M., {et~al.} 2009, \apj, 690, 1236

\bibitem[{{Iwasawa} {et~al.}(2009){Iwasawa}, {Sanders}, {Evans}, {Mazzarella},
  {Armus}, \& {Surace}}]{iwasawa09}
{Iwasawa}, K., {Sanders}, D.~B., {Evans}, A.~S., {et~al.} 2009, \apjl, 695,
  L103

\bibitem[{{Knapen} {et~al.}(2000){Knapen}, {Shlosman}, \&
  {Peletier}}]{knapen00}
{Knapen}, J.~H., {Shlosman}, I., \& {Peletier}, R.~F. 2000, \apj, 529, 93

\bibitem[{{Kocevski} {et~al.}(2012){Kocevski}, {Faber}, {Mozena}, {Koekemoer},
  {Nandra}, {Rangel}, {Laird}, {Brusa}, {Wuyts}, {Trump}, {Koo}, {Somerville},
  {Bell}, {Lotz}, {Alexander}, {Bournaud}, {Conselice}, {Dahlen}, {Dekel},
  {Donley}, {Dunlop}, {Finoguenov}, {Georgakakis}, {Giavalisco}, {Guo},
  {Grogin}, {Hathi}, {Juneau}, {Kartaltepe}, {Lucas}, {McGrath}, {McIntosh},
  {Mobasher}, {Robaina}, {Rosario}, {Straughn}, {van der Wel}, \&
  {Villforth}}]{kocevski12}
{Kocevski}, D.~D., {Faber}, S.~M., {Mozena}, M., {et~al.} 2012, \apj, 744, 148

\bibitem[{{Koekemoer} {et~al.}(2007){Koekemoer}, {Aussel}, {Calzetti}, {Capak},
  {Giavalisco}, {Kneib}, {Leauthaud}, {Le F{\`e}vre}, {McCracken}, {Massey},
  {Mobasher}, {Rhodes}, {Scoville}, \& {Shopbell}}]{cosmos_acs}
{Koekemoer}, A.~M., {Aussel}, H., {Calzetti}, D., {et~al.} 2007, \apjs, 172,
  196

\bibitem[{{Kormendy} \& {Kennicutt}(2004)}]{kormendy&kennicutt04}
{Kormendy}, J., \& {Kennicutt}, Jr., R.~C. 2004, \araa, 42, 603

\bibitem[{{Kraljic} {et~al.}(2012){Kraljic}, {Bournaud}, \&
  {Martig}}]{kraljic12}
{Kraljic}, K., {Bournaud}, F., \& {Martig}, M. 2012, \apj, 757, 60

\bibitem[{{Laine} {et~al.}(2002){Laine}, {Shlosman}, {Knapen}, \&
  {Peletier}}]{laine02}
{Laine}, S., {Shlosman}, I., {Knapen}, J.~H., \& {Peletier}, R.~F. 2002, \apj,
  567, 97

\bibitem[{{Leauthaud} {et~al.}(2007){Leauthaud}, {Massey}, {Kneib}, {Rhodes},
  {Johnston}, {Capak}, {Heymans}, {Ellis}, {Koekemoer}, {Le F{\`e}vre},
  {Mellier}, {R{\'e}fr{\'e}gier}, {Robin}, {Scoville}, {Tasca}, {Taylor}, \&
  {Van Waerbeke}}]{cosmos_leauthaud07}
{Leauthaud}, A., {Massey}, R., {Kneib}, J., {et~al.} 2007, \apjs, 172, 219

\bibitem[{{Lehmer} {et~al.}(2010){Lehmer}, {Alexander}, {Bauer}, {Brandt},
  {Goulding}, {Jenkins}, {Ptak}, \& {Roberts}}]{lehmer10}
{Lehmer}, B.~D., {Alexander}, D.~M., {Bauer}, F.~E., {et~al.} 2010, \apj, 724,
  559

\bibitem[{{Lira} {et~al.}(2002){Lira}, {Ward}, {Zezas}, {Alonso-Herrero}, \&
  {Ueno}}]{lira02}
{Lira}, P., {Ward}, M., {Zezas}, A., {Alonso-Herrero}, A., \& {Ueno}, S. 2002,
  \mnras, 330, 259

\bibitem[{{Marinova} \& {Jogee}(2007)}]{marinova&jogee07}
{Marinova}, I., \& {Jogee}, S. 2007, \apj, 659, 1176

\bibitem[{{Martin}(1995)}]{martin95}
{Martin}, P. 1995, \aj, 109, 2428

\bibitem[{{Martini}(2004)}]{martini04_lifetime}
{Martini}, P. 2004, in Coevolution of Black Holes and Galaxies, ed. L.~C.~Ho
  (Cambridge: Cambridge University Press), 169

\bibitem[{{Martini} {et~al.}(2001){Martini}, {Pogge}, {Ravindranath}, \&
  {An}}]{martini01}
{Martini}, P., {Pogge}, R.~W., {Ravindranath}, S., \& {An}, J.~H. 2001, \apj,
  562, 139

\bibitem[{{Martini} {et~al.}(2003){Martini}, {Regan}, {Mulchaey}, \&
  {Pogge}}]{martini03b}
{Martini}, P., {Regan}, M.~W., {Mulchaey}, J.~S., \& {Pogge}, R.~W. 2003, \apj,
  589, 774

\bibitem[{{Melvin} {et~al.}(2014){Melvin}, {Masters}, {Lintott}, {Nichol},
  {Simmons}, {Bamford}, {Casteels}, {Cheung}, {Edmondson}, {Fortson},
  {Schawinski}, {Skibba}, {Smith}, \& {Willett}}]{melvin14}
{Melvin}, T., {Masters}, K., {Lintott}, C., {et~al.} 2014, \mnras, 438, 2882

\bibitem[{{Menci} {et~al.}(2014){Menci}, {Gatti}, {Fiore}, \&
  {Lamastra}}]{menci14}
{Menci}, N., {Gatti}, M., {Fiore}, F., \& {Lamastra}, A. 2014, \aap, 569, A37

\bibitem[{{Men{\'e}ndez-Delmestre} {et~al.}(2007){Men{\'e}ndez-Delmestre},
  {Sheth}, {Schinnerer}, {Jarrett}, \& {Scoville}}]{menendez07}
{Men{\'e}ndez-Delmestre}, K., {Sheth}, K., {Schinnerer}, E., {Jarrett}, T.~H.,
  \& {Scoville}, N.~Z. 2007, \apj, 657, 790

\bibitem[{{Mullaney} {et~al.}(2012){Mullaney}, {Pannella}, {Daddi},
  {Alexander}, {Elbaz}, {Hickox}, {Bournaud}, {Altieri}, {Aussel}, {Coia},
  {Dannerbauer}, {Dasyra}, {Dickinson}, {Hwang}, {Kartaltepe}, {Leiton},
  {Magdis}, {Magnelli}, {Popesso}, {Valtchanov}, {Bauer}, {Brandt}, {Del Moro},
  {Hanish}, {Ivison}, {Juneau}, {Luo}, {Lutz}, {Sargent}, {Scott}, \&
  {Xue}}]{mullaney12a}
{Mullaney}, J.~R., {Pannella}, M., {Daddi}, E., {et~al.} 2012, \mnras, 419, 95

\bibitem[{{Murata} {et~al.}(2014){Murata}, {Kajisawa}, {Taniguchi},
  {Kobayashi}, {Shioya}, {Capak}, {Ilbert}, {Koekemoer}, {Salvato}, \&
  {Scoville}}]{murata14}
{Murata}, K.~L., {Kajisawa}, M., {Taniguchi}, Y., {et~al.} 2014, \apj, 786, 15

\bibitem[{{Peng} {et~al.}(2010){Peng}, {Ho}, {Impey}, \& {Rix}}]{galfit10}
{Peng}, C.~Y., {Ho}, L.~C., {Impey}, C.~D., \& {Rix}, H. 2010, \aj, 139, 2097

\bibitem[{{Ramos Almeida} {et~al.}(2012){Ramos Almeida}, {Bessiere},
  {Tadhunter}, {P{\'e}rez-Gonz{\'a}lez}, {Barro}, {Inskip}, {Morganti}, {Holt},
  \& {Dicken}}]{ramosalmeida12}
{Ramos Almeida}, C., {Bessiere}, P.~S., {Tadhunter}, C.~N., {et~al.} 2012,
  \mnras, 419, 687

\bibitem[{{Regan} \& {Mulchaey}(1999)}]{regan99a}
{Regan}, M.~W., \& {Mulchaey}, J.~S. 1999, \aj, 117, 2676

\bibitem[{{Regan} {et~al.}(1999){Regan}, {Sheth}, \& {Vogel}}]{regan99b}
{Regan}, M.~W., {Sheth}, K., \& {Vogel}, S.~N. 1999, \apj, 526, 97

\bibitem[{{Rosario} {et~al.}(2013){Rosario}, {Mozena}, {Wuyts}, {Nandra},
  {Koekemoer}, {McGrath}, {Hathi}, {Dekel}, {Donley}, {Dunlop}, {Faber},
  {Ferguson}, {Giavalisco}, {Grogin}, {Guo}, {Kocevski}, {Koo}, {Laird},
  {Newman}, {Rangel}, \& {Somerville}}]{rosario13}
{Rosario}, D.~J., {Mozena}, M., {Wuyts}, S., {et~al.} 2013, \apj, 763, 59

\bibitem[{{Sakamoto} {et~al.}(1999){Sakamoto}, {Okumura}, {Ishizuki}, \&
  {Scoville}}]{sakamoto99}
{Sakamoto}, K., {Okumura}, S.~K., {Ishizuki}, S., \& {Scoville}, N.~Z. 1999,
  \apj, 525, 691

\bibitem[{{Salvato} {et~al.}(2011){Salvato}, {Ilbert}, {Hasinger}, {Rau},
  {Civano}, {Zamorani}, {Brusa}, {Elvis}, {Vignali}, {Aussel}, {Comastri},
  {Fiore}, {Le Floc'h}, {Mainieri}, {Bardelli}, {Bolzonella}, {Bongiorno},
  {Capak}, {Caputi}, {Cappelluti}, {Carollo}, {Contini}, {Garilli}, {Iovino},
  {Fotopoulou}, {Fruscione}, {Gilli}, {Halliday}, {Kneib}, {Kakazu},
  {Kartaltepe}, {Koekemoer}, {Kovac}, {Ideue}, {Ikeda}, {Impey}, {Le Fevre},
  {Lamareille}, {Lanzuisi}, {Le Borgne}, {Le Brun}, {Lilly}, {Maier},
  {Manohar}, {Masters}, {McCracken}, {Messias}, {Mignoli}, {Mobasher}, {Nagao},
  {Pello}, {Puccetti}, {Perez-Montero}, {Renzini}, {Sargent}, {Sanders},
  {Scodeggio}, {Scoville}, {Shopbell}, {Silvermann}, {Taniguchi}, {Tasca},
  {Tresse}, {Trump}, \& {Zucca}}]{cosmos_salvato11}
{Salvato}, M., {Ilbert}, O., {Hasinger}, G., {et~al.} 2011, \apj, 742, 61

\bibitem[{{Sanders} {et~al.}(1988){Sanders}, {Soifer}, {Elias}, {Madore},
  {Matthews}, {Neugebauer}, \& {Scoville}}]{sanders88a}
{Sanders}, D.~B., {Soifer}, B.~T., {Elias}, J.~H., {et~al.} 1988, \apj, 325, 74

\bibitem[{{Santini} {et~al.}(2012){Santini}, {Rosario}, {Shao}, {Lutz},
  {Maiolino}, {Alexander}, {Altieri}, {Andreani}, {Aussel}, {Bauer}, {Berta},
  {Bongiovanni}, {Brandt}, {Brusa}, {Cepa}, {Cimatti}, {Daddi}, {Elbaz},
  {Fontana}, {F{\"o}rster Schreiber}, {Genzel}, {Grazian}, {Le Floc'h},
  {Magnelli}, {Mainieri}, {Nordon}, {P{\'e}rez Garcia}, {Poglitsch}, {Popesso},
  {Pozzi}, {Riguccini}, {Rodighiero}, {Salvato}, {Sanchez-Portal}, {Sturm},
  {Tacconi}, {Valtchanov}, \& {Wuyts}}]{santini12}
{Santini}, P., {Rosario}, D.~J., {Shao}, L., {et~al.} 2012, \aap, 540, A109

\bibitem[{{Schawinski} {et~al.}(2011){Schawinski}, {Treister}, {Urry},
  {Cardamone}, {Simmons}, \& {Yi}}]{schawinski11}
{Schawinski}, K., {Treister}, E., {Urry}, C.~M., {et~al.} 2011, \apjl, 727, L31

\bibitem[{{Scoville} {et~al.}(2007){Scoville}, {Aussel}, {Brusa}, {Capak},
  {Carollo}, {Elvis}, {Giavalisco}, {Guzzo}, {Hasinger}, {Impey}, {Kneib},
  {LeFevre}, {Lilly}, {Mobasher}, {Renzini}, {Rich}, {Sanders}, {Schinnerer},
  {Schminovich}, {Shopbell}, {Taniguchi}, \& {Tyson}}]{cosmos}
{Scoville}, N., {Aussel}, H., {Brusa}, M., {et~al.} 2007, \apjs, 172, 1

\bibitem[{{Sheth} {et~al.}(2012){Sheth}, {Melbourne}, {Elmegreen}, {Elmegreen},
  {Athanassoula}, {Abraham}, \& {Weiner}}]{sheth12}
{Sheth}, K., {Melbourne}, J., {Elmegreen}, D.~M., {et~al.} 2012, \apj, 758, 136

\bibitem[{{Sheth} {et~al.}(2000){Sheth}, {Regan}, {Vogel}, \&
  {Teuben}}]{sheth00}
{Sheth}, K., {Regan}, M.~W., {Vogel}, S.~N., \& {Teuben}, P.~J. 2000, \apj,
  532, 221

\bibitem[{{Sheth} {et~al.}(2005){Sheth}, {Vogel}, {Regan}, {Thornley}, \&
  {Teuben}}]{sheth05}
{Sheth}, K., {Vogel}, S.~N., {Regan}, M.~W., {Thornley}, M.~D., \& {Teuben},
  P.~J. 2005, \apj, 632, 217

\bibitem[{{Sheth} {et~al.}(2008){Sheth}, {Elmegreen}, {Elmegreen}, {Capak},
  {Abraham}, {Athanassoula}, {Ellis}, {Mobasher}, {Salvato}, {Schinnerer},
  {Scoville}, {Spalsbury}, {Strubbe}, {Carollo}, {Rich}, \& {West}}]{sheth08}
{Sheth}, K., {Elmegreen}, D.~M., {Elmegreen}, B.~G., {et~al.} 2008, \apj, 675,
  1141

\bibitem[{{Shlosman} {et~al.}(1989){Shlosman}, {Frank}, \&
  {Begelman}}]{shlosman89}
{Shlosman}, I., {Frank}, J., \& {Begelman}, M.~C. 1989, \nat, 338, 45

\bibitem[{{Tacconi} {et~al.}(2010){Tacconi}, {Genzel}, {Neri}, {Cox}, {Cooper},
  {Shapiro}, {Bolatto}, {Bouch{\'e}}, {Bournaud}, {Burkert}, {Combes},
  {Comerford}, {Davis}, {Schreiber}, {Garcia-Burillo}, {Gracia-Carpio}, {Lutz},
  {Naab}, {Omont}, {Shapley}, {Sternberg}, \& {Weiner}}]{tacconi10}
{Tacconi}, L.~J., {Genzel}, R., {Neri}, R., {et~al.} 2010, \nat, 463, 781

\bibitem[{{Treister} {et~al.}(2012){Treister}, {Schawinski}, {Urry}, \&
  {Simmons}}]{treister12}
{Treister}, E., {Schawinski}, K., {Urry}, C.~M., \& {Simmons}, B.~D. 2012,
  \apjl, 758, L39

\bibitem[{{Trump} {et~al.}(2014){Trump}, {Barro}, {Juneau}, {Weiner}, {Luo},
  {Brammer}, {Bell}, {Brandt}, {Dekel}, {Guo}, {Hopkins}, {Koo}, {Kocevski},
  {McIntosh}, {Momcheva}, {Faber}, {Ferguson}, {Grogin}, {Kartaltepe},
  {Koekemoer}, {Lotz}, {Maseda}, {Mozena}, {Nandra}, {Rosario}, \&
  {Zeimann}}]{trump14}
{Trump}, J.~R., {Barro}, G., {Juneau}, S., {et~al.} 2014, ArXiv:1407.7525,
  arXiv:1407.7525

\bibitem[{{Ueda} {et~al.}(2014){Ueda}, {Akiyama}, {Hasinger}, {Miyaji}, \&
  {Watson}}]{ueda14}
{Ueda}, Y., {Akiyama}, M., {Hasinger}, G., {Miyaji}, T., \& {Watson}, M.~G.
  2014, \apj, 786, 104

\end{thebibliography}

\newcommand{\noopsort}[1]{}

\end{document}